%% file: Main.tex
\documentclass[review, 12pt, authoryear]{elsarticle}
\usepackage[english]{babel}
\usepackage{lineno}%,hyperref}
\graphicspath{Figures/}
\usepackage{graphicx}
\usepackage{booktabs} % For formal tables
\usepackage{enumerate}
\usepackage{subcaption}
\usepackage{times}
\usepackage{ulem}
\usepackage[colorinlistoftodos]{todonotes}
\usepackage{natbib}
\usepackage{amsmath}
\usepackage{url}
\usepackage{array}
\usepackage{acro}
\usepackage[document]{ragged2e}
\input{Acronym.tex}
\usepackage[left=1in, right=1in, top=1in, bottom=1in]{geometry}
\newcolumntype{L}[1]{>{\raggedright\let\newline\\\arraybackslash\hspace{0pt}}m{#1}}
\newcolumntype{C}[1]{>{\centering\let\newline\\\arraybackslash\hspace{0pt}}m{#1}}
\newcolumntype{R}[1]{>{\raggedleft\let\newline\\\arraybackslash\hspace{0pt}}m{#1}}
\makeatletter
\newcommand\xLongLeftRightArrow[2][]{%
	\ext@arrow 0099{\LongLeftRightArrowfill@}{#1}{#2}}
\def\LongLeftRightArrowfill@{%
	\arrowfill@\Leftarrow\Relbar\Rightarrow}
\makeatother
\usepackage{tablefootnote}
\usepackage{multirow}

\journal{International Journal of Human Computer Studies}

\begin{document}
	%\acuseall
	\acsetup{first-style=long}
	\begin{frontmatter}
\title{Data-Driven Vibrotactile Rendering of Digital Buttons on Touchscreens}
\author{Bushra Sadia, Senem Ezgi Emgin, T. Metin Sezgin, Cagatay Basdogan\corref{CorrespondingAuthor}} 
\address{College of Engineering, Ko\c{c} University, 34450, Istanbul, Turkey}

\cortext[CorrespondingAuthor]{Email: cbasdogan@ku.edu.tr (Cagatay Basdogan), \\ College of Engineering,  Ko\c{c} University, 34450, Istanbul, Turkey.}
\begin{abstract}
Interaction with physical buttons is an essential part of our daily routine. We use buttons daily to turn lights on, to call an elevator, to ring a doorbell, or even to turn on our mobile devices. Buttons have  distinct response characteristics and are easily activated by touch. However, there is limited tactile feedback available for their digital counterparts displayed on touchscreens. Although mobile phones incorporate low-cost vibration motors to enhance touch-based interactions, it is not possible to generate complex tactile effects on touchscreens. It is also difficult to relate the limited vibrotactile feedback generated by these motors to different types of physical buttons. In this study, we focus on creating vibrotactile feedback on a touchscreen that simulates the feeling of physical buttons using piezo actuators attached to it. We first recorded and analyzed the force, acceleration, and voltage data from twelve participants interacting with three different physical buttons: latch, toggle, and push buttons. Then, a button-specific vibrotactile stimulus was generated for each button based on the recorded data. Finally, we conducted a three-alternative forced choice (3AFC) experiment with twenty participants to explore whether the resultant stimulus is distinct and realistic. In our experiment, participants were able to match the three digital buttons with their physical counterparts with a success rate of 83\%. In addition, we harvested seven adjective pairs from the participants expressing their perceptual feeling of pressing the physical buttons. All twenty participants rated the degree of their subjective feelings associated with each adjective for all the physical and digital buttons investigated in this study. Our statistical analysis showed that there exist at least three adjective pairs for which participants have rated two out of three digital buttons similar to their physical counterparts.
\end{abstract}

\acuseall
\end{frontmatter}
\section{Introduction}
\justify
Touchscreens are an integral part of our mobile phones, tablets, laptops, ATMs, tabletops, vending machines, electronic kiosks, and car navigation systems. These interactive screens have replaced the physical buttons with the digital ones. However, the lack of sophisticated tactile feedback in digital buttons results in a decrease in user experience quality and even task performance (\citealp{c7, c6,kaaresoja2016latency}).

Pressing a physical button is a basic interaction method for everyday tasks such as turning on a light, ringing a doorbell, or activating a laptop/mobile phone. Digital buttons can be used to send an email, write a message, dial a phone number, or type a digital keyword to search the Web. When a digital button is activated by pressing, typically visual (such as change of appearance, shape, and color) or audio (such as button click sound) feedback or both are displayed to inform users of their actions. Compared to vision and sound, haptics has been utilized less as sensory feedback for digital buttons \citep{lee2009performance}.

To display haptic feedback for interactions with digital buttons, researchers have utilized various types of electromechanical actuators, such as voice coils, vibration motors (\citealp{nashel2003tactile, c5, c2, c3, c8}) and piezoelectric actuators (\citealp{c1, pakkanen2010comparison, c9, c10, c12, c13, c14, ma2015haptic, c20}). 

\cite{c3} investigated the cross-modal congruency between the visual and audio/tactile feedback for digital buttons of mobile touchscreens. They used Eccentric Resonance Mass (ERM) and piezoelectric actuators to create four different vibrotactile stimuli for displaying eight visually different digital buttons in terms of shape, size, and height. Their study revealed that users could successfully relate between the visual and audio/tactile properties of buttons. They also compared a standard touchscreen with the one displaying tactile feedback and concluded that the addition of tactile feedback significantly reduces the errors in finger-based text entry (\citealp{c2}). \cite{c4} investigated the characteristics of digital buttons displayed by piezoelectric actuators and vibration motors (ERMs) on touch screens to identify the type of tactile click that is perceived as most pleasant to the finger. They showed that digital buttons with tactile feedback are superior to the ones without tactile feedback, regardless of the technology used to create the tactility. Their results also showed that piezoelectric actuators created more pleasant feedback compared with vibration motors.

\cite{c8} focused on displaying tactile feedback on mobile devices when a digital button is pressed on the screen. They presented seventy-two different vibrotactile stimulus to the subjects by varying amplitude, duration, carrier signal, envelope function, and type of actuators. These waveforms are analogous to the typical acceleration profiles that are observed during contacts of a pen/stylus with a rigid surface (\citealp{okamura2001reality}), which can be modeled using an exponentially decaying sinusoidal function. They suggested that, for pleasant button clicks, short rise time as well as short duration are preferable. \cite{c9} designed and evaluated a set of simulated key clicks for key-less mobile devices equipped with piezoelectric actuators. They concluded that key clicks are perceived as ``crisp" (``dull") at the stimulation frequency of 500 Hz (125 Hz). \cite{c10} investigated the preference in type of tactile feedback displayed by piezo-actuated digital buttons under different time delays and vibration durations. Their results showed that, it is possible to create either favorable or significantly less favorable button click feedback by varying delay and duration parameters within a relatively short time window. They also concluded that the signal for tactile feedback requires careful design and control of the duration parameter. Haptic feedback can also be displayed during the process of pushing down a button  (\citealp{kim2013haptic}). This type of feedback mimics the force-displacement curve of a mechanical button to further enhance the perceived realism of a digital button.

A button-like click feeling can also be generated via squeeze film effect (\citealp{c1, c19, c20}) by varying the friction coefficient between the user's finger and touchscreen.

The earlier studies have mainly focused on displaying ``pleasant'' button click sensations using vibration cues, but not the realistic feeling. However, creating vibrotactile stimuli for digital buttons with distinct features based on the measured data originating from their mechanical counterparts has not been considered before. Researchers have mainly used this approach for realistic rendering of viscoelastic materials using a force feedback device (\citealp{ c22}), realistic textures displayed by a voice coil actuator (\citealp{c21, c23, c24}), and electrostatic actuation on touchscreens (\citealp{c25}).

In this study, we focus on three different types of physical buttons to display their digital counterparts on a touchscreen: \textit{Latch Button}, \textit{Toggle Button}, and \textit{Push Button}. For that purpose, we recorded force, acceleration, and voltage data for activation state from those buttons while twelve participants were pressing on them. Our experimental data reveals that these three buttons have distinct features in term of (a) the magnitude of normal force required for activation, (b) the activation state with respect to the instant of voltage change, and (c) the resulting mechanical vibrations (acceleration) of the button when pressed. We mapped the recorded acceleration signal of each physical button to a voltage signal that actuates the piezo patches of an interactive multi-touch table (\citealp{emgin2018haptable}) for displaying their digital counterparts on its touch surface. We then conducted two experimental studies to investigate how these digital buttons were perceived by the user. Our first study was a matching experiment, where twenty participants participated in a three-alternative forced choice (3AFC) experiment to match the digital buttons with their physical counterparts. Our second study investigated the perceptual feelings arising from pressing three physical and digital buttons. A very recent study by \cite{liu2018perceptual} investigated the perceptual dimensions associated with manual key clicks. However, it was limited to only different types of push buttons. In our second experiment, we first collected a set of adjectives from the participants that describes their perceptual feelings of pressing three physical buttons. Then, participants rated the degree of feelings associated with each adjective for the physical and digital buttons investigated in this study.

\section{Methodology}
\begin{figure}[h!]
	\centering
	\includegraphics[width=0.5\textwidth]{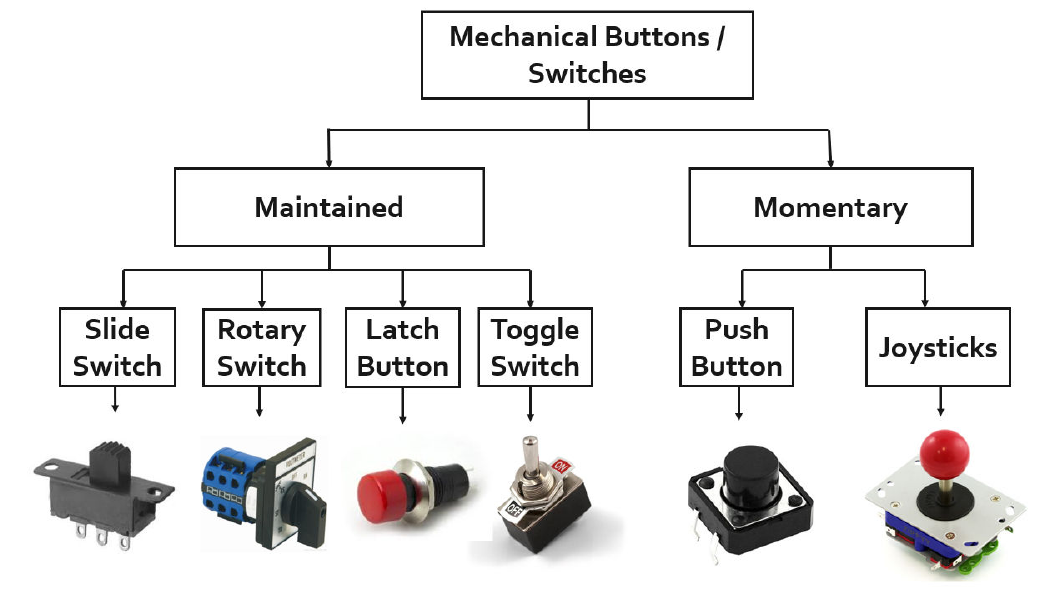}
	\caption{Physical buttons/switches can be divided into two main groups. Our study utilizes \textit{Latch Button} and \textit{Toggle Button} from maintained group, and \textit{Push Button} from momentary group.}\label{fig1}
	%  %\vspace{-0.5cm}
\end{figure}

Mechanical buttons and switches are designed to enable or disable current flow in electric circuits. A button must be triggered by an external force to change its state, which can be achieved by different hand gestures (pushing, rotating, sliding, rocking, pulling). The response of each button depends on its unique properties, and accordingly, buttons can be grouped under two main categories (Fig. \ref{fig1}) (\citealp{c34}): 

\begin{enumerate}[(a)]
	\item Maintained buttons/switches stay in one state until they are triggered. Toggle, rotary, latch, and slide buttons are examples of maintained buttons.
	\item Momentary buttons/switches stay active as long as pressure is maintained on the switch. Examples of momentary buttons/switches include push buttons and joysticks. 
\end{enumerate}

\begin{figure}[h!]
	\centering
	
	\includegraphics[width=0.4\textwidth]{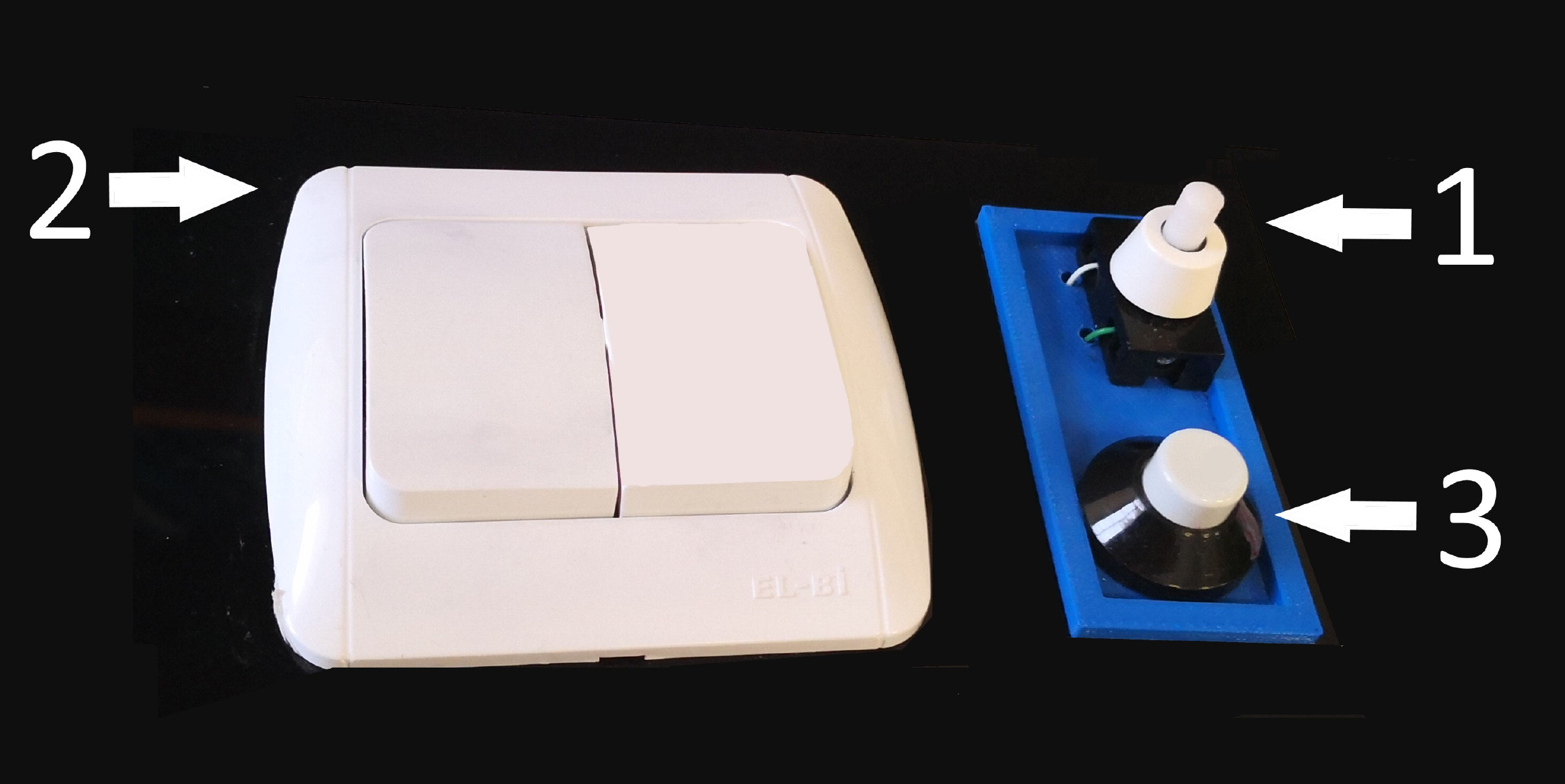}
	\vspace{0.2cm}
	\caption{Physical buttons used during data collection and in our second experiment: \textit{Latch Button} (1), \textit{Toggle Button} (2), and \textit{Push Button} (3).}\label{fig2}
\end{figure}

We chose three buttons that we interact frequently in our daily lives: two maintained buttons (\textit{Latch Button} and light button as \textit{Toggle Button}), and one momentary button (\textit{Push Button}) (Fig. \ref{fig2}). The acceleration profile (mechanical vibrations that occur when a physical button is pressed) of buckling (pressing a button) and restitution (releasing a button) events are very similar, but they occur in opposite directions (\citealp{c1}). For rendering purpose,  we focus only on the buckling event of these three buttons.  

\subsection{Experimental Setup}
We used a custom designed tabletop, HapTable (\citealp{emgin2018haptable}), that registers and recognizes touch gestures using rear diffused illumination (\citealp{c16}) and is capable of providing appropriate visual and haptic feedback for the recognized gestures. 

The interaction surface of HapTable is evenly illuminated with wide angle infrared LEDs (50-Module IR Kit, Environmental Lights). When a user touches this surface, light is reflected from contact points and captured by an infrared camera (Eye 3, PlayStation). In this study, we added an additional infrared camera (Eye 3, PlayStation) underneath the touch surface  at a shorter distance to the interaction surface in order to capture  higher-resolution images of the finger contact area as the user presses a digital button. 

To generate mechanical vibrations on the surface, HapTable uses a sound card, a high voltage amplifier (E413.D2, Physik Instrumente, Gain: 50), and a solid state relay (Yocto - MaxiCoupler, Yoctopuce). The sound card generates haptic signals, which are amplified by the high voltage amplifier and sent to the solid state relay. The outputs of the solid state relay are connected to individual piezo patches, which can be actuated in less than ten milliseconds. In order to obtain the vibrational characteristics of the touch surface of HapTable, a linear sine sweep voltage signal, varying in frequency from 0 to 625 Hz, was applied to piezo patches using a signal generator. Then, a Laser Doppler Vibrometer (LDV, PDV-100, Polytec) was used to measure the out-of-plane vibrations at 84 grid points on the touch surface, and a signal analyzer (NetDB, 01dB-Metravib) was used to record and analyze these signals. We constructed five vibration maps of the touch surface for each actuation configuration of piezo actuators: when each piezo actuator was excited individually (PA, PB, PC, and PD), and when all piezo actuators were excited together. For further details on vibration maps, readers are referred to \cite{emgin2018haptable}.

\begin{figure}[h!]
	\centering
	\includegraphics[width=0.45\textwidth]{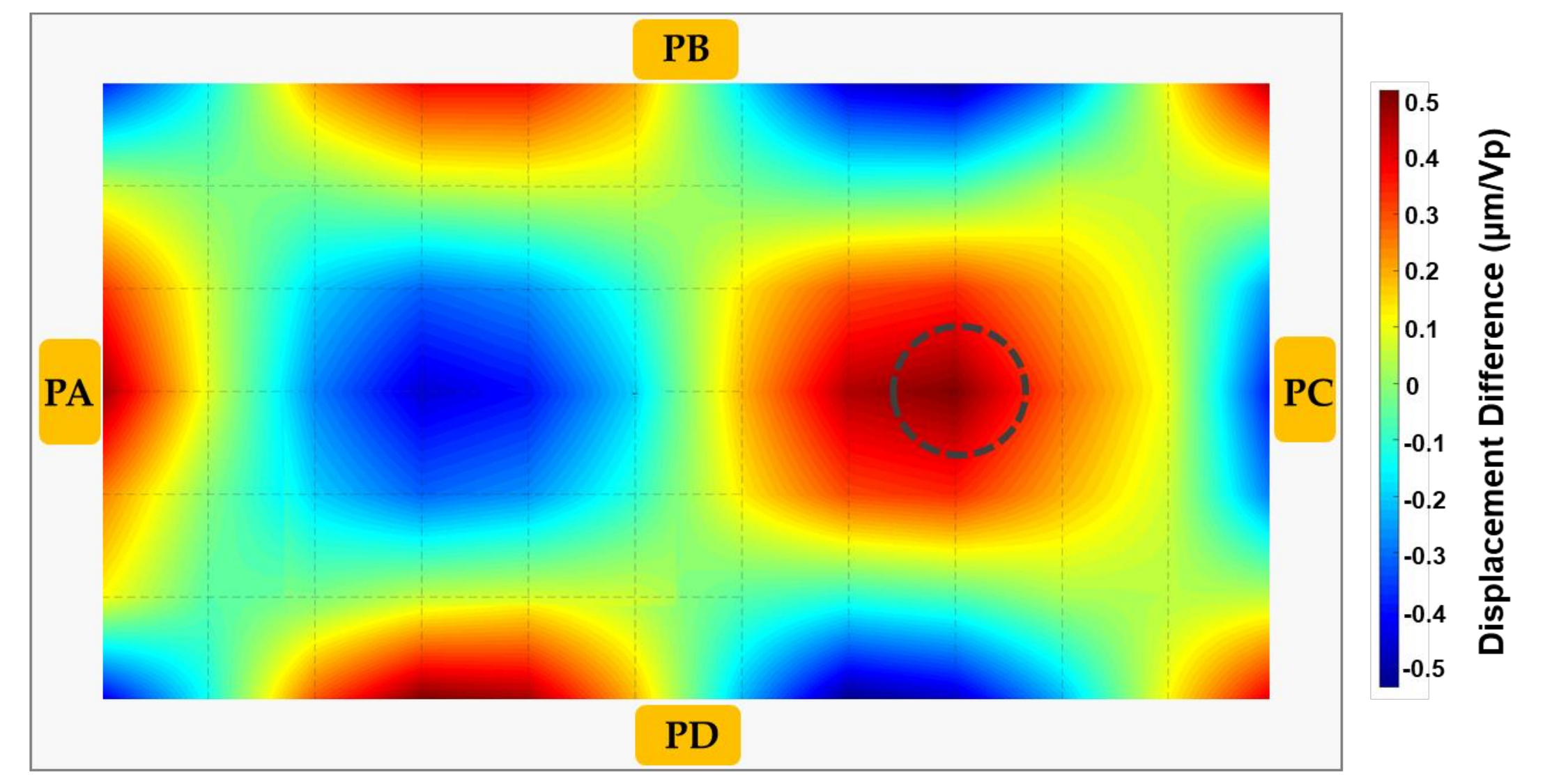}
	\vspace{0.2cm}
	\caption{Vibration map of the touch surface when piezo patch A (PA) is excited at 263.5 Hz. The digital buttons are rendered at the center of dashed circle. }\label{fig3}
\end{figure}

The vibration maps generated for these five cases revealed that the highest vibration amplitude with the largest area of interaction was achieved at a resonance frequency of 263.5 Hz using piezo patch A (PA) at the region shown in Fig. \ref{fig3}. This resonance frequency is desirable for our application since human tactile sensitivity to vibrotactile stimulus is highest around 250Hz (\citealp{c31}). 

\subsection{Data Collection} \label{sec21}
\begin{figure}[h!]
	\centering
	\includegraphics[width=0.45\textwidth]{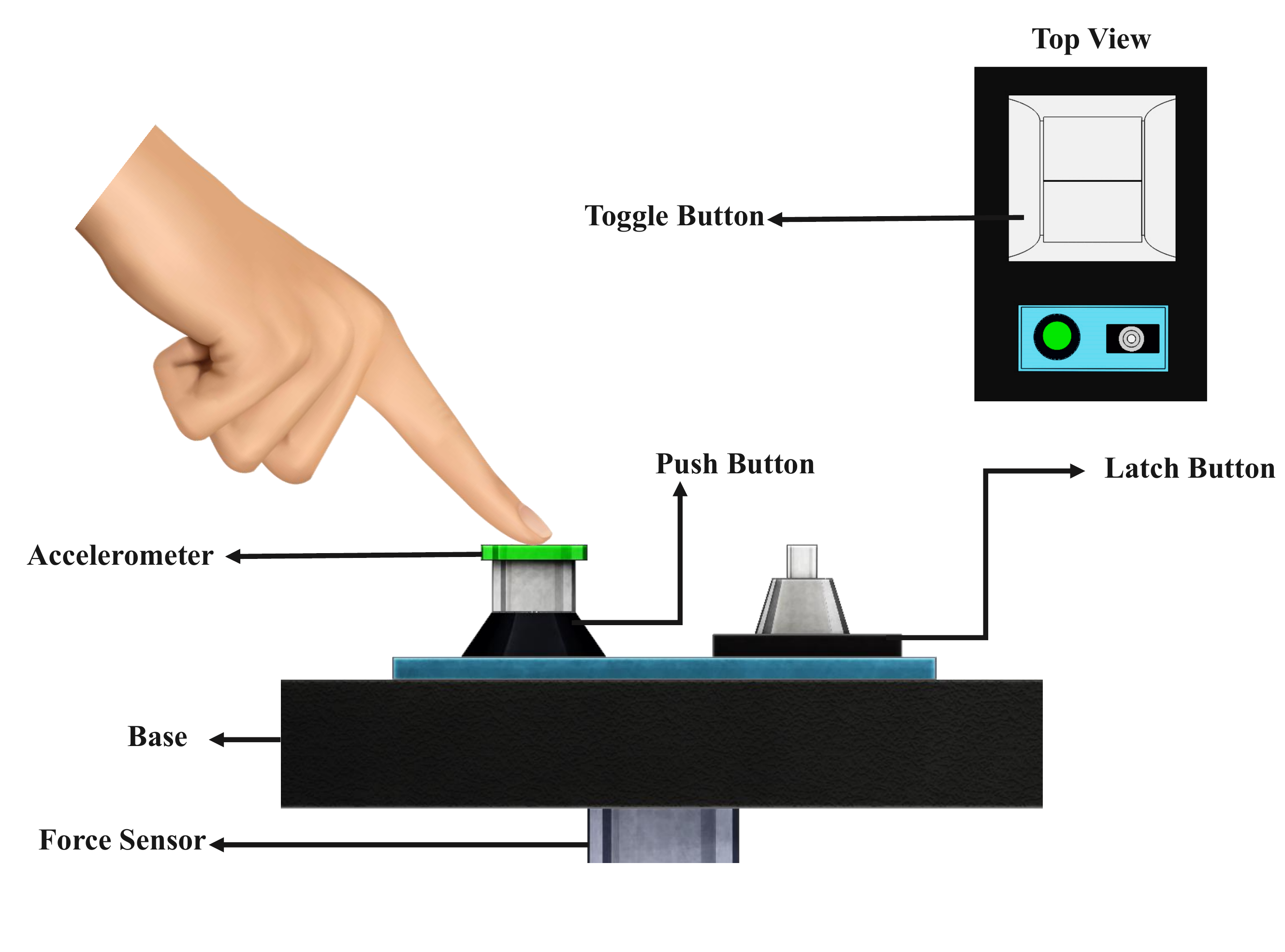}
	\vspace{0.2cm}
	\caption{Illustration of the experimental setup used for data collection.}\label{fig4} 
\end{figure}

To measure the force applied by the user on each physical button, its vertical acceleration, and activation state in the form of a step change in voltage, we built a box enclosing all three buttons (Fig. \ref{fig4}). A three-axis accelerometer (GY-61, Analog Inc.) was attached on top of each button  to measure its vertical acceleration and a force sensor (mini-40, ATI Industrial Automation) was placed beneath the box to measure the normal force applied by the user. Two data acquisition cards (PCI-6034E and PCI-6321E, National Instruments) were used to record the force, acceleration, and  voltage data  simultaneously at 5K samples per seconds. The box was immobilized on a rigid table to reduce the electrical noise due to cabling and undesired external vibrations.

Twelve participants (eight males and four females with an average age of 31.12 $\pm$ 5.82 years), participated in the data collection experiment. Each participant read, and then signed a consent form approved by the Ethical Committee for Human Participants of the Ko\c{c} University before the experiment. Participants were asked to press each button ten times using their index finger. During this interaction, we recorded the force applied to button, button's acceleration, and  activation state in the form of a step change in the voltage signal. Collecting experimental data from each participant took approximately fifteen minutes.

\subsection{Data Processing}
\begin{figure}[h!]
	\centering
	\includegraphics[width=0.5\textwidth]{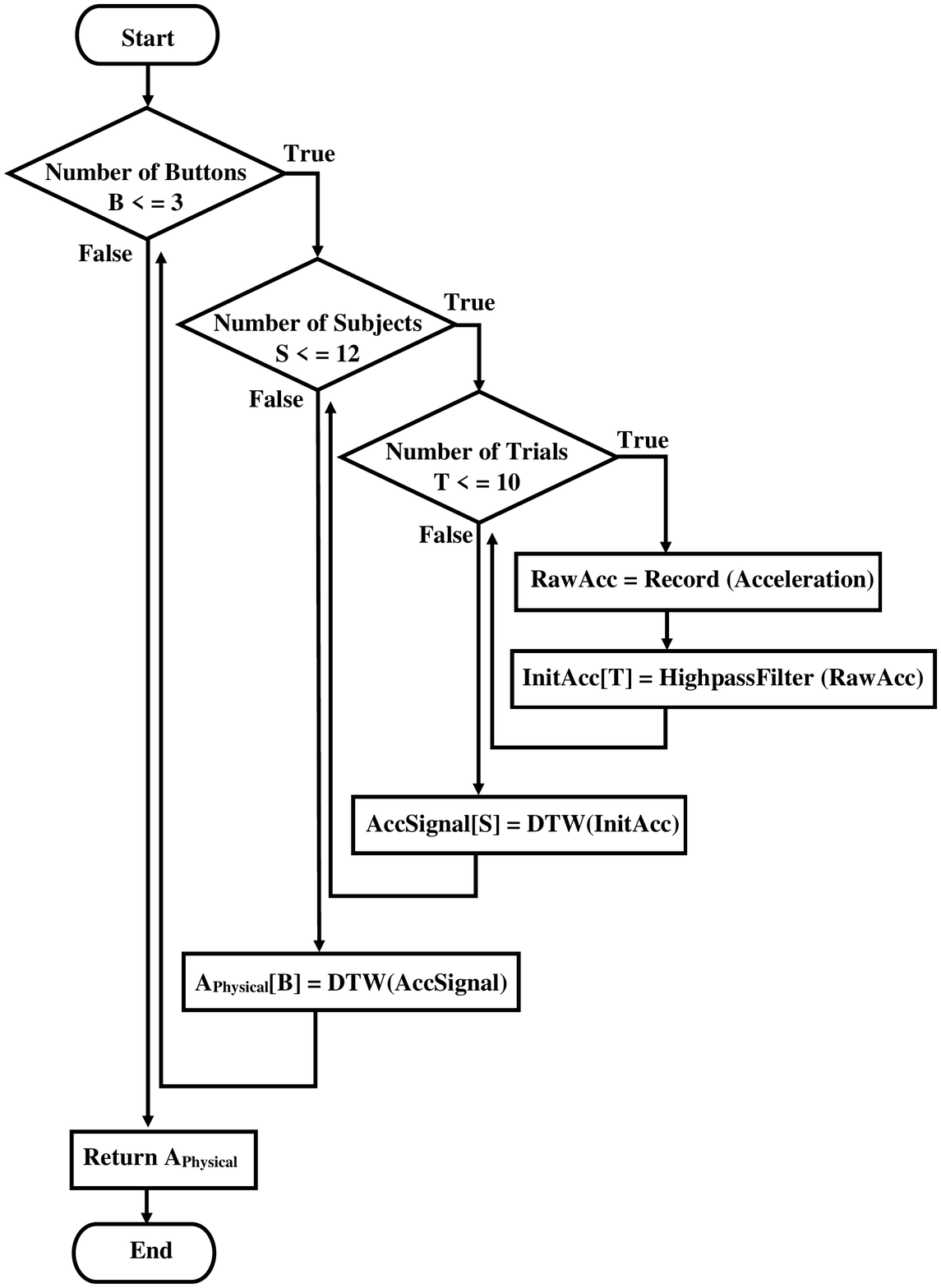}
	\vspace{0.2cm}
	\caption{Flowchart showing the steps involved in collecting and processing acceleration data. In this chart, \textbf{RawAcc} is the recorded acceleration, \textbf{InitAcc} is the filtered acceleration, \textbf{AccSignal} is the representative signal of a subject and $A_{physical}$ is an array containing the representative acceleration signal of all three physical buttons.}\label{fig5}
\end{figure}

Before processing the accelerometer data, we applied a high pass filter with a cutoff frequency of 10 Hz to eliminate an undesired influence of gravitational forces, sensor drift, and user's hand movements (\citealp{c21}). Then, we applied dynamic time warping (DTW) algorithm (\citealp{c15}) to the acceleration signals to find a representative signal for each button. This algorithm aligns two time-dependent signals by minimizing the total distance between them. The flowchart used for collecting and processing of acceleration data is shown in Fig. \ref{fig5}.

\begin{figure}[h!]
	\centering
	\includegraphics[width=0.7\textwidth]{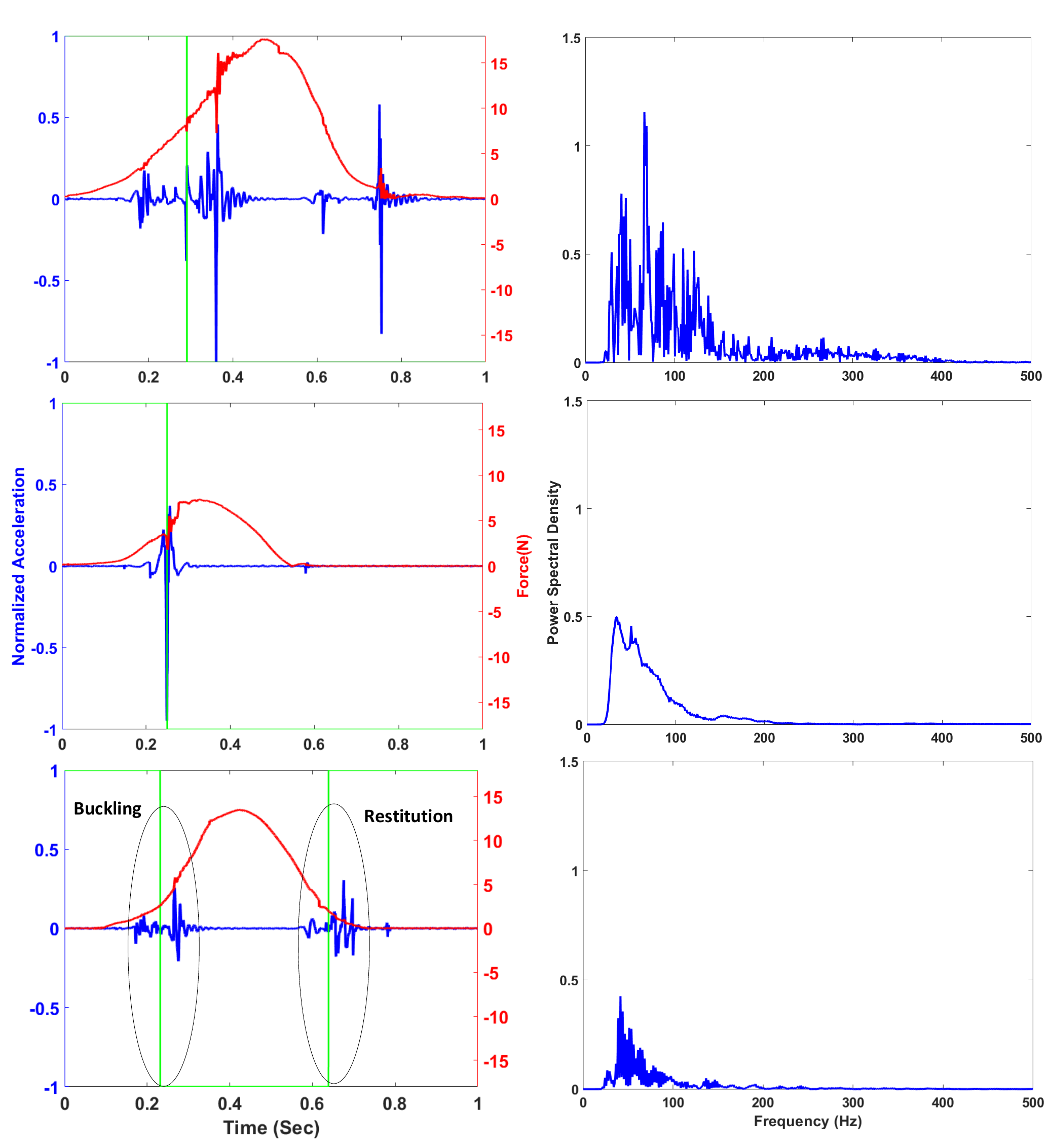}
	\vspace{-0.2cm}
	\caption{Normalized acceleration (blue), force (red), and  activation state (green) are shown for \textit{Latch Button} (top), \textit{Toggle Button} (middle), and \textit{Push Button} (bottom) along with the frequency spectrum of the acceleration signals on the right column. The buckling event is shown for \textit{Latch} and \textit{Toggle Buttons} while both events are shown for \textit{Push Button} since it is a momentary type of button.
	}\label{fig6}
\end{figure}

\begin{figure}[h!]
	\centering
	\vspace{-0.2cm}
	\includegraphics[width=0.45\textwidth]{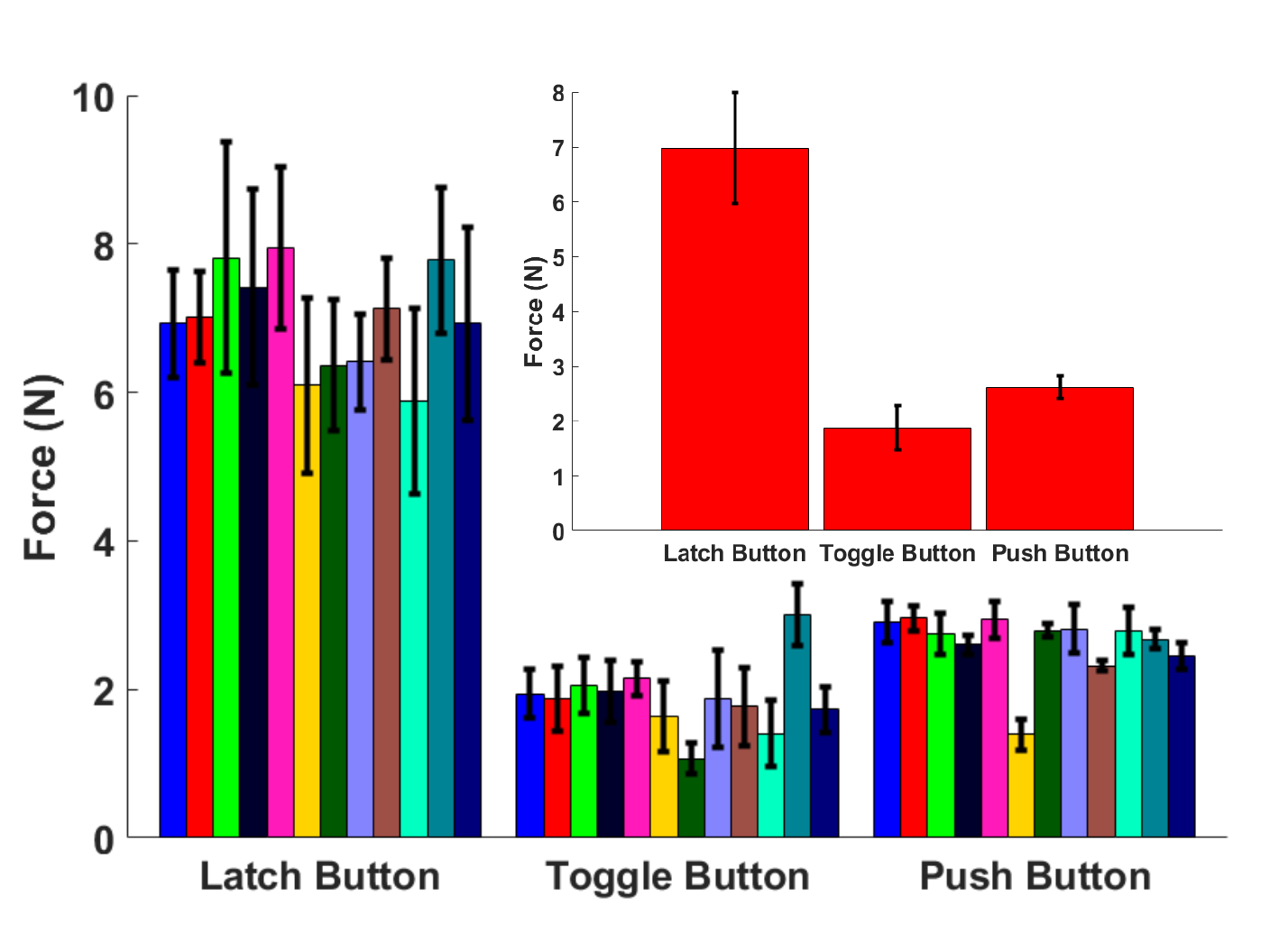}
	\vspace{0.2cm}
	\caption{ Mean and standard deviation of force magnitudes for each participant and all together (upper right). The force data was acquired at the instant of button activation.}\label{fig7}
\end{figure}

Typical acceleration, force, and activation state profiles recorded from each physical button and the power spectral densities of acceleration signals obtained by Fast-Fourier transform are shown in Fig. \ref{fig6}. As shown in Fig. \ref{fig6}, the measured acceleration signals have distinct forms in the frequency domain. Fig. \ref{fig7} reports the magnitude of the force applied by each participant to activate each physical button. Fig. \ref{fig7} also shows that the participants applied the largest force to activate \ac{b1} while \ac{b2} required the smallest amount of force to activate.

In order to construct a vibrotactile waveform for each digital button, we performed amplitude modulation on the acceleration signal recorded from its physical counterpart with the resonant frequency of the touch surface of HapTable (263.5 Hz) using Equation \ref{eq6}: 

\begin{equation}
A_{digital,i}(t) = A_{physical,i}(t) sin(2\pi f_ct)
\label{eq6}
\end{equation}
where $i$ represents the button type,  $A_{physical,i}(t)$, and $A_{digital,i}(t)$ are the recorded and synthesized acceleration signals of each button and $f_c$ is the carrier frequency (263.5 Hz). 

\subsection{Transfer Function Estimation}
\begin{figure}[!h]
	\centering
	\begin{center}
		\includegraphics[width=0.3\textwidth]{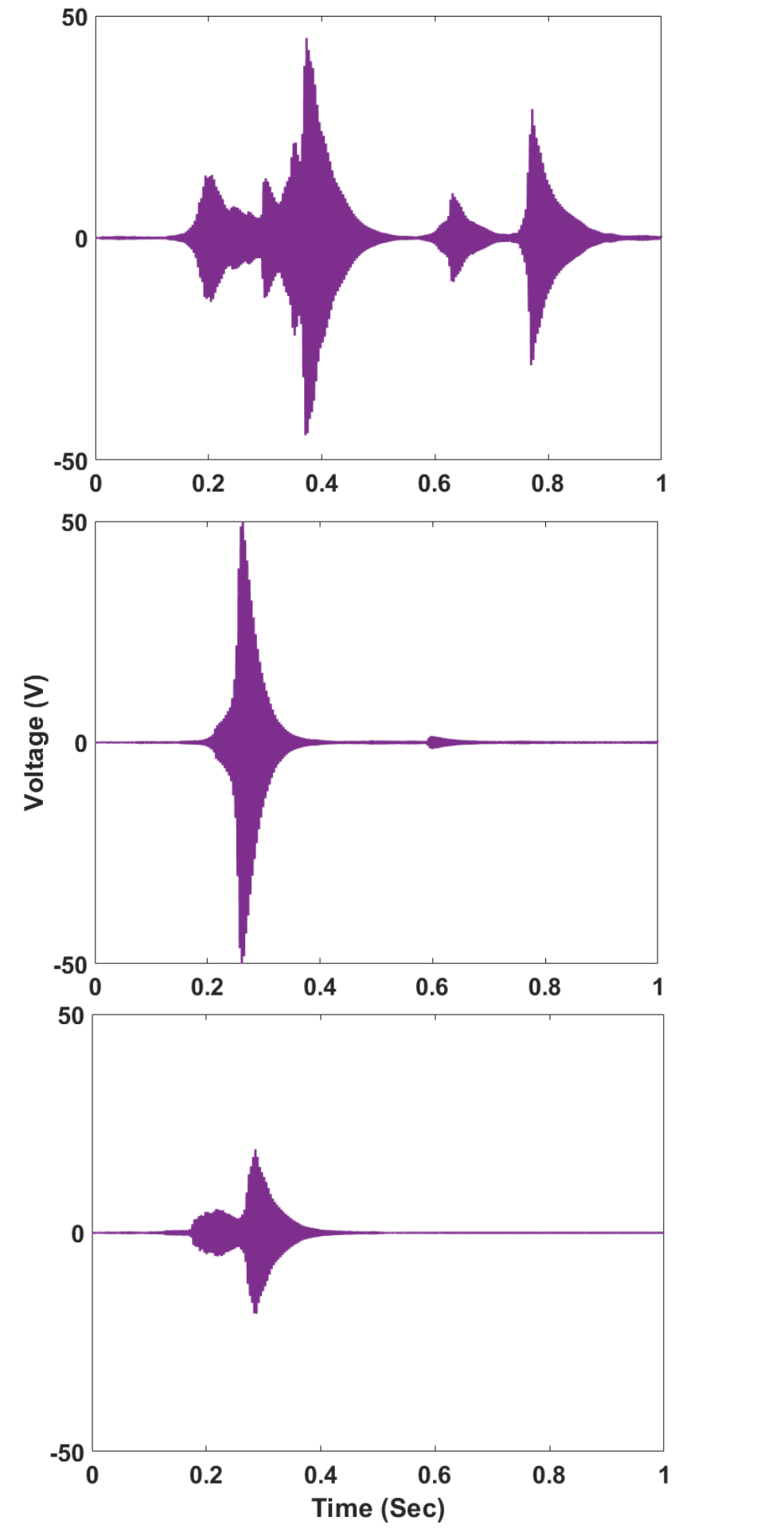}
		\vspace{1em}
		\caption{Voltage signals applied to the piezo patch A for digital \textit{Latch Button} (top), \textit{Toggle Button} (middle), and \textit{Push Button} (bottom). The input voltage is 100Vpp.}\label{fig8}
	\end{center}
\end{figure}

In order to preserve the spectrum of synthesized waveform $A_{digital}(t)$, we estimated the transfer function between the voltage applied to the piezo patches and the acceleration generated on the touch surface of HapTable using the Frequency Response Function (FRF) recorded for piezo patch A (PA) between the voltage and vibration amplitude, at the point where the digital buttons are displayed to the user (Fig. \ref{fig3}). Readers are referred to \cite{emgin2018haptable} for further details on the computation of such an FRF. We differentiated the FRF twice to get the acceleration/voltage, which was originally measured in the form of displacement/voltage. We then inverted the FRF (voltage/acceleration) and fit a  transfer function  using  $``tfest"$ function of MATLAB. Finally, we  inputted   this transfer function and the synthesized acceleration signal $A_{digital,i}(t)$ to $``lsim"$ function of MATLAB to obtain the actuation voltage signal. The voltage signals obtained by this method for all  digital buttons are shown in Fig. \ref{fig8}.

\begin{figure}[!h]
	\centering
	\begin{center}
		\includegraphics[width=0.7\textwidth]{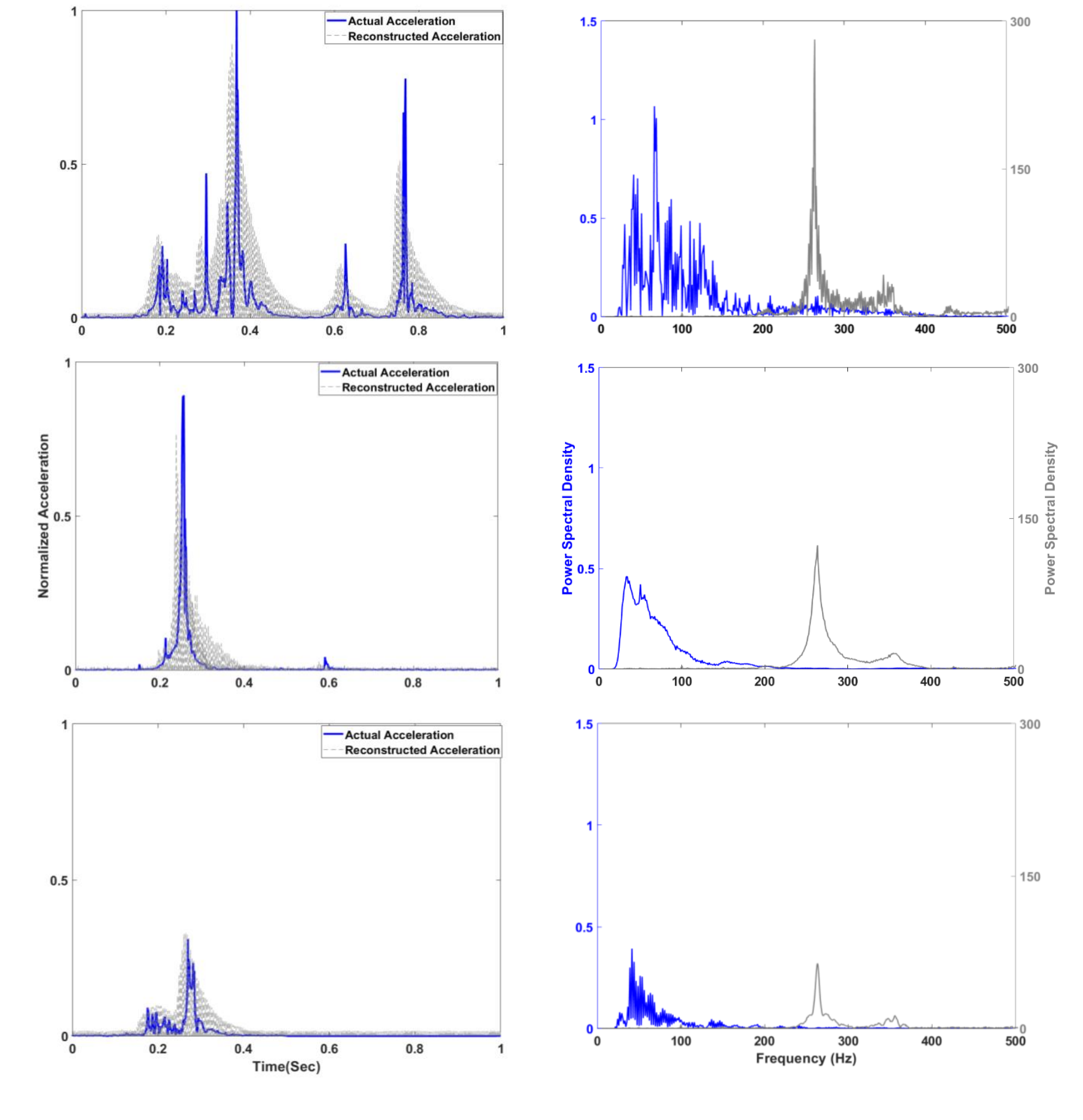}
		\vspace{0.2cm}
		\caption{Comparison of the reconstructed acceleration signals (gray) with the recorded acceleration signals (blue) for \textit{Latch Button} (top), \textit{Toggle Button} (middle), and \textit{Push Button} (bottom) along with the frequency spectrum of the recorded acceleration signals (blue) and reconstructed acceleration signals (gray) on the right column.}\label{fig9}
	\end{center}
\end{figure}

We applied the voltage signals to the piezo patch A (PA in Fig. \ref{fig3}) and measured the corresponding vibrational velocities at the interaction point on the touch surface (where the digital buttons are displayed to the participant) by LDV. We then differentiated the measured velocities to get the acceleration profiles. Finally, we compared the   reconstructed acceleration signals with the actual  acceleration signals recorded from physical buttons in Fig. \ref{fig9}. We observed a good match between the actual and reconstructed acceleration signals in the time domain. Obviously, they did not match in the frequency domain since the frequency spectrum of the reconstructed acceleration signals were dominated by the resonance frequency of the HapTable. 

\subsection{Haptic Rendering of Digital Buttons based on Finger Contact Area} \label{sec22}
Force is an important factor in actuating a physical button (\citealp{colton2007reality,c32}). Some buttons are actuated by lightly tapping on the keys (such as the \textit{Toggle Button} in our case) while others need higher actuation force (such as the \textit{Latch Button} in our case). Typically, touch surfaces are not equipped with force sensors. Therefore, we cannot directly use force information to actuate the digital buttons on mobile phones and tablets.  

\begin{figure}[h!]
	\centering
	\vspace{-0.2cm}
	\includegraphics[width=0.45\textwidth]{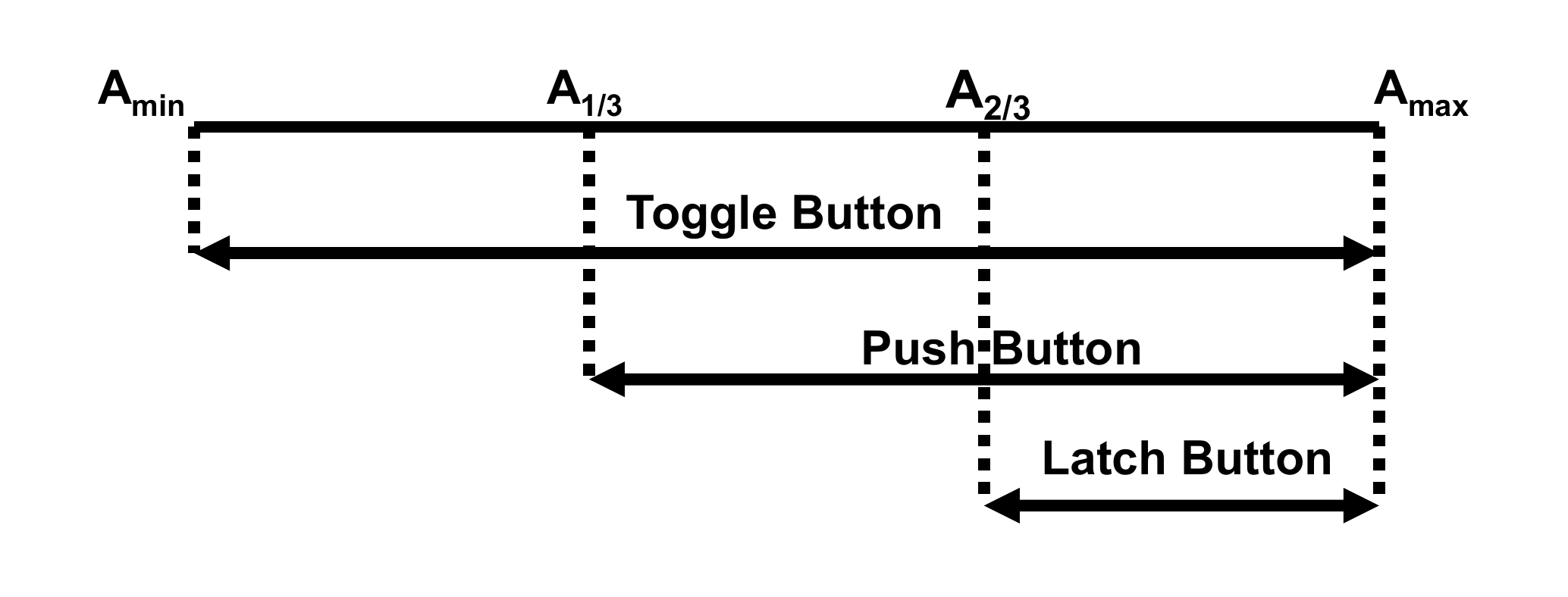}
	\vspace{0.2cm}
	\caption{Conditions for activating digital buttons based on finger contact area.}\label{fig10}
	%  %\vspace{-0.5cm}
\end{figure}

Existing literature shows that the apparent finger contact area grows proportionally with applied normal force (\citealp{c27, c26}). Therefore, we related the mean normal force required to activate  physical buttons (see the upper right corner in Fig. \ref{fig7}) with the finger contact area of the participants to activate digital buttons.  Hence, \ac{b1} required the largest area of finger contact while \ac{b2} required the smallest area of finger contact. For implementation purposes, we simply defined three conditions to activate each digital button as depicted in Fig. \ref{fig10}, where $A_{min}$ and $A_{max}$ represent the contact areas corresponding to low and high finger pressure. The digital button is activated when the participant applies sufficient pressure to the touch surface with her/his finger such that the instantaneous contact area exceeds the pre-determined threshold values (Fig. \ref{fig10}). 

Obviously, the mapping between normal force and finger contact area is not linear when Hertz contact model is considered. However, our visualization table (HapTable) is not equipped with a force/pressure sensor. Therefore, it was not possible to activate the digital buttons based on the recorded force values of participants. Moreover, we did not measure the finger contact area of the participants simultaneously with the force applied to the physical buttons in our experimental setup (Fig. \ref{fig4}) during our recording session (Section 2.2). Also, we did not control their finger orientation while they pressed the physical buttons, which affects the size of the contact area. For those reasons, in our study, we measured the finger contact area of each participant using an IR camera and then use a simple linear model, based on the recorded force data from physical buttons, to activate the digital buttons.

\section{Experiment-I: Matching Experiment}
In this experiment, we asked participants to match digital buttons with their physical counterparts by pressing the digital button on the interaction surface of HapTable and relate it with one of the three physical buttons.  

\subsection{Participants}
Twenty participants (12 males and 8 females with an average age of 29.12 $\pm$ 6.42 years) participated in this experiment. Sixteen participants were right-handed and the rest were left-handed. All of them were graduate students and everyday users of mobile phones. The participants read and signed a consent form before the experiment. The form was approved by the Ethical Committee for Human Participants of Ko\c{c} University.

\subsection{Experimental Design and Procedure} \label{sec32}
A $2\times 2$ latin square design was used to investigate if the participants' matching performance was improved after learning the corresponding stimulus for each physical button (\citealp{foehrenbach2009tactile, ma2015haptic}). The participants were randomly assigned to one of the four experimental groups. Each participant took part in two experimental sessions conducted on two consecutive days. Each session involved four steps as shown in Fig. \ref{fig11} below:

\begin{figure}[h!]
	\centering
	\includegraphics[width=0.5\textwidth]{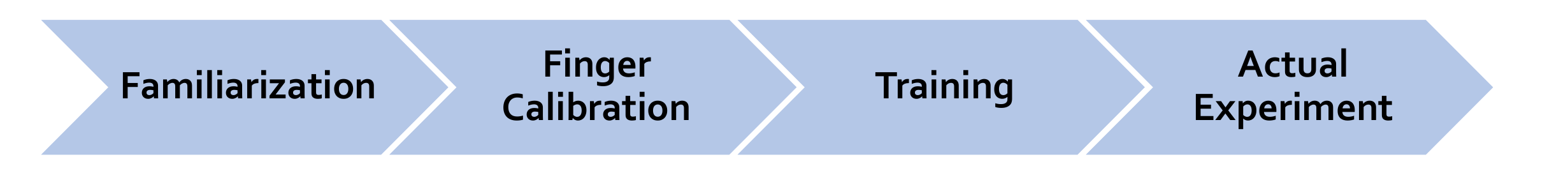}
	\vspace{0.2cm}
	\caption{Four steps involved in the matching experiment.}\label{fig11} 
\end{figure}

\textbf{Familiarization:} Participants pressed each of the three physical buttons, ten times to familiarize themselves with the haptic feedback resulting from pressing these buttons. They were blindfolded to prevent any visual bias but instructed to pay attention to the tactile cues while they were interacting with physical buttons.

\textbf{Finger Calibration:} As reported in the earlier studies (\citealp{c28,c29}), humans have different finger contact areas for the same normal force. To display personalized haptic feedback to each participant in our study, we measured the participants' finger contact area before the actual experiment while they were pressing on the touch surface, as if they were activating a digital button, under two different contact conditions: (a) low ($A_{min}$) and (b) high ($A_{max}$) finger pressure. Participants were asked to press the touch surface five times for each contact condition. The mean values for $A_{min}$ and $A_{max}$ were utilized in activating digital buttons based on the conditions set in Fig. \ref{fig10}. 

\textbf{Training:} This step was designed to familiarize participants with the experimental setup, and the haptic feedback displayed for the digital buttons on HapTable. During this step, participants were allowed to ask questions. Each participant completed nine trials (3 repetition x 3 digital buttons). In each trial, participants received haptic feedback according to their contact area and button type and asked to match it with one of the three physical buttons. We provided confirmation to Group-II about the correctness of their selected choices during the training step on Day-I, but not on Day-II. Group-I received confirmation about the correctness of their choice during this step on Day-II, but not on Day-I (Table \ref{tab1}).

\begin{table}[h!]
	\centering
	\caption{Latin square design used for Experiment-I.}
	\label{tab1}
	\centering
	\renewcommand{\arraystretch}{1.5}
	\begin{tabular}{|c|c|c|ll}
		\cline{1-3}
		\textbf{}       & \textbf{Day-I} & \textbf{Day-II} &  &  \\ \cline{1-3}
		\textbf{Group-I} & {No Confirmation}   & {Confirmation}  &  &  \\ \cline{1-3}
		\textbf{Group-II} & {Confirmation}  & {No Confirmation}   &  &  \\ \cline{1-3}
	\end{tabular}
	\renewcommand{\arraystretch}{1}
\end{table}

\textbf{Actual Experiment:} There were thirty trials in the actual experiment (10 repetitions x 3 digital buttons). Participants were not allowed to ask any questions. They wore noise-canceling headphones to eliminate audio cues. The task was to press the digital button on the HapTable to feel the vibrotactile signal and relate it with one of the three physical buttons they experienced earlier in the familiarization step. 

\section{Experiment-II: Subjective Rating Experiment}
Experiment-II aimed to investigate how effectively we were able to render the haptic feeling of physical buttons in the digital domain. The experiment consisted of two steps. In the first step, we collected adjectives from the participants through a survey, describing their tactile sensations of the three physical buttons. In the second step, participants experimented with the physical and digital buttons and rated their subjective experiences via those adjectives. 

\subsection{Collecting Adjectives}
The participants were asked to write down the adjectives that could be associated with the tactile sensation of all three physical buttons. We sorted the most frequently appeared adjectives in their list and selected the top nine adjective pairs with opposite meanings (Table \ref{tab2}). We removed the stiff/spring adjective pair because our current setup cannot render the stiffness/springiness of a button. 

\begin{table}[]
	\caption{List of nine adjective pairs used for adjective ratings in Experiment-II.}
	\label{tab2}
	\centering
	\renewcommand{\arraystretch}{1.5}
	\begin{tabular}{|c|ccc|}
		\hline
		Pair No. & \multicolumn{3}{c|}{Adjectives}                  \\ \hline 
		1        & Unpleasant    & $\xLongLeftRightArrow{\hspace*{1.4cm}} $ & Pleasant    \\
		2        & Uncomfortable & $\xLongLeftRightArrow{\hspace*{1.4cm}}$ & Comfortable \\
		3        & Unclear       & $\xLongLeftRightArrow{\hspace*{1.4cm}}$ & Clear       \\
		4        & Unstable      & $\xLongLeftRightArrow{\hspace*{1.4cm}}$ & Stable      \\
		5        & Delayed       & $\xLongLeftRightArrow{\hspace*{1.4cm}}$ & Quick       \\
		6        & Unreliable    & $\xLongLeftRightArrow{\hspace*{1.4cm}}$ & Reliable    \\
		7        & Rough         & $\xLongLeftRightArrow{\hspace*{1.4cm}}$ & Smooth    \\
		8*         & Hard        & $\xLongLeftRightArrow{\hspace*{1.4cm}}$ & Soft        \\ 
		9*         & Stiff       & $\xLongLeftRightArrow{\hspace*{1.4cm}}$ & Springy	    \\\hline 
		\multicolumn{4}{l}{\footnotesize{*Adjective pair was removed after the pilot study.}}
	\end{tabular}
	\renewcommand{\arraystretch}{1}
\end{table}

After a pilot study conducted with ten new participants, we also removed the hard/soft adjective pair as the participants reported difficulty in attributing hardness/softness to the digital buttons. Hence, we ended up with seven adjective pairs. 
 
\subsection{Subjective Evaluation}
\begin{figure}[h!]
	\centering
	%\vspace{-0.5cm}
	\includegraphics[width=0.5\textwidth]{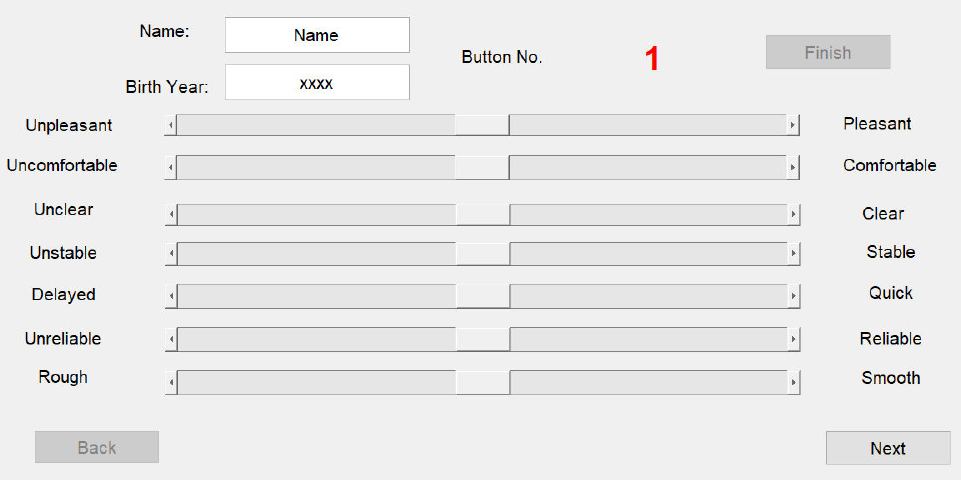}
	\vspace{0.2cm}
	\caption{The user interface for the adjective ratings.}\label{fig12}
\end{figure}

In this step, the participants rated their tactile sensations of physical and digital buttons using the seven adjective pairs shown in Table \ref{tab2}. We designed a simple GUI to collect adjective ratings for each of the three physical and digital buttons (Fig. \ref{fig12}). 
Participants were allowed to experiment with each button as many times as they wanted. Then, they were asked to manipulate the slider bar of each adjective pair to rate their subjective feelings. All slider bars were centered at the beginning of evaluation for each button. On average, the experiment took thirty minutes for each participant. 

\section{Results}
\subsection{Experiment-I}
\begin{figure*}[!htp]
	%\fulltextwidth
	\centering
	\begin{subfigure}[b]{0.45\textwidth}
		\centering
		\caption{ }	\label{fig13a}
		\includegraphics[width=0.94\linewidth]{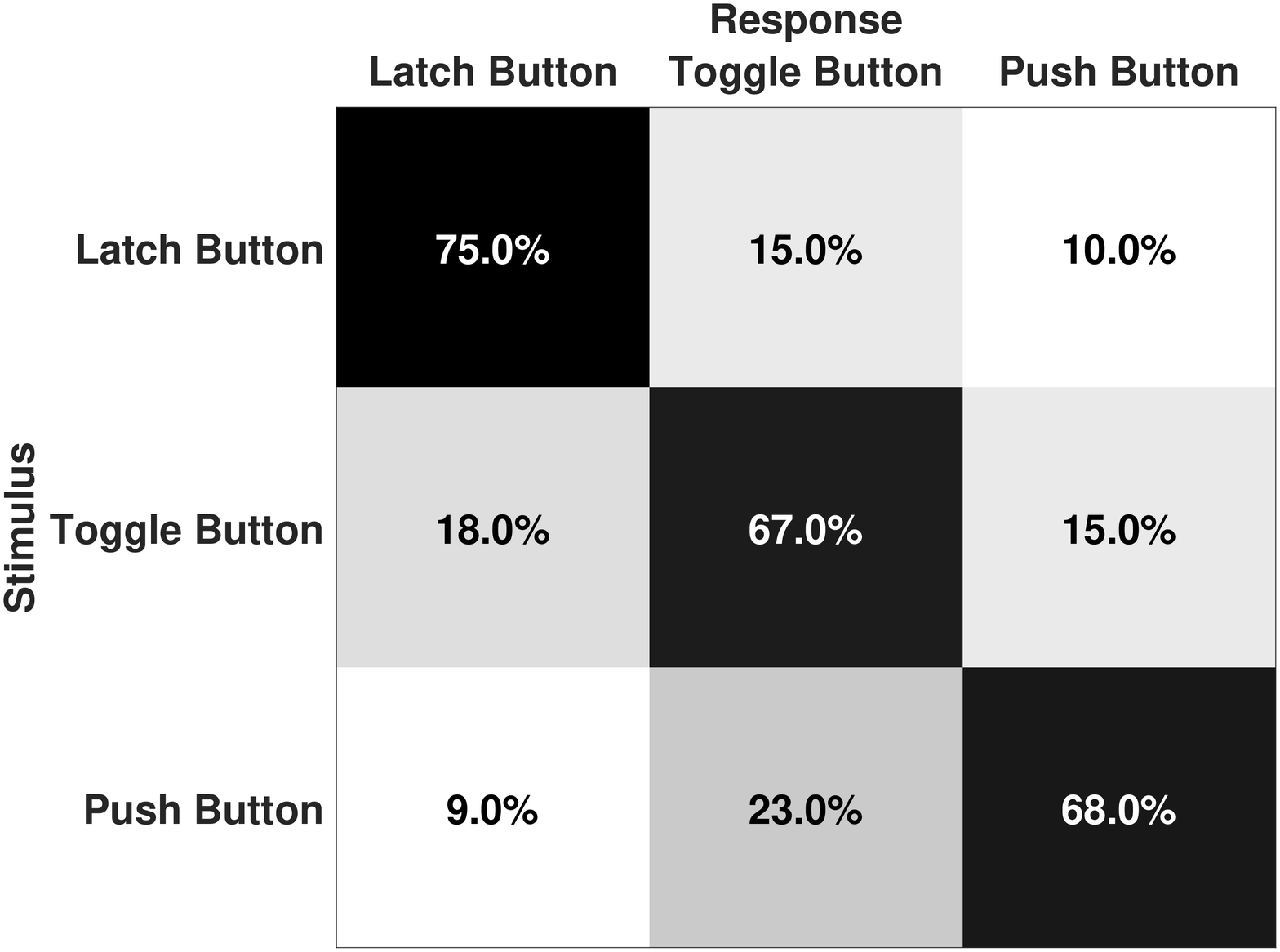}
		
	\end{subfigure}%
	%	\vspace{0.5} 
	\vspace{1em}
	\begin{subfigure}[b]{0.45\textwidth}
		\centering
		\caption{ } \label{fig13b}
		\includegraphics[width=1\linewidth]{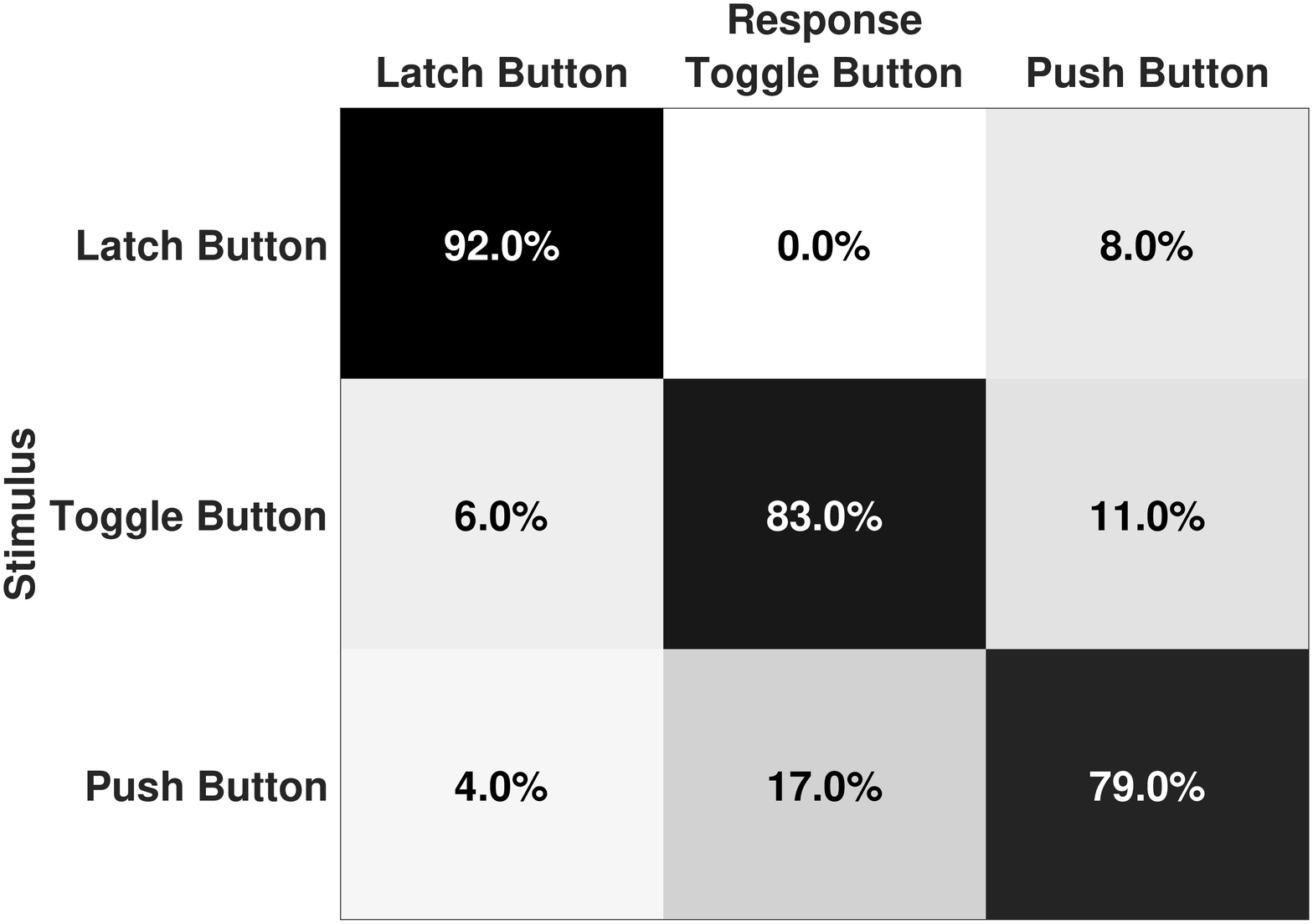}
		
	\end{subfigure}
	
	\begin{subfigure}[b]{0.45\textwidth}
		\centering
		\caption{ }	\label{fig13c}
		\includegraphics[width=1\linewidth]{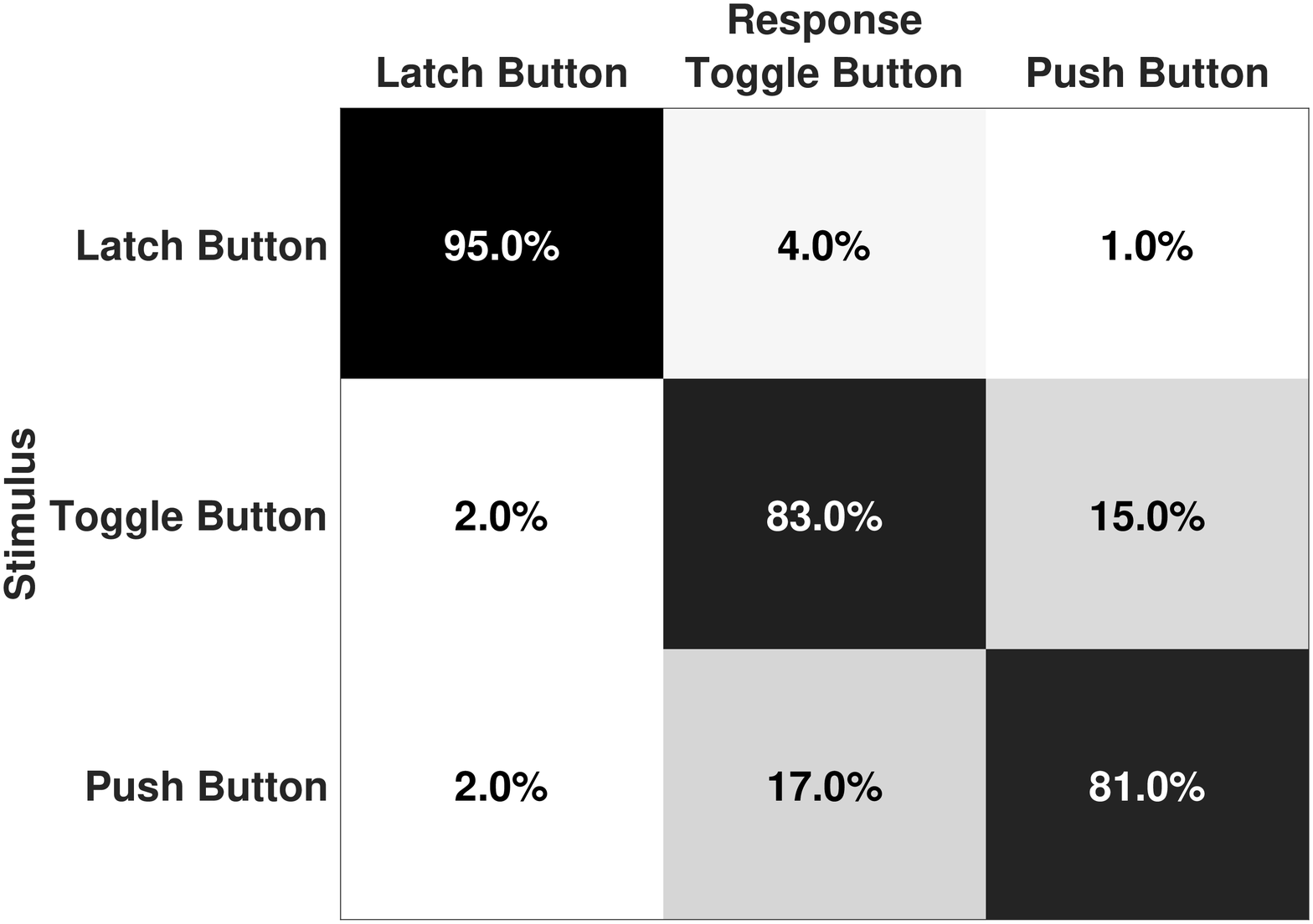}
		
	\end{subfigure}%
	%	\vspace{0.5} 
	\vspace{1em}
	\begin{subfigure}[b]{0.45\textwidth}
		\centering
		\caption{ } \label{fig13d}
		\includegraphics[width=1\linewidth]{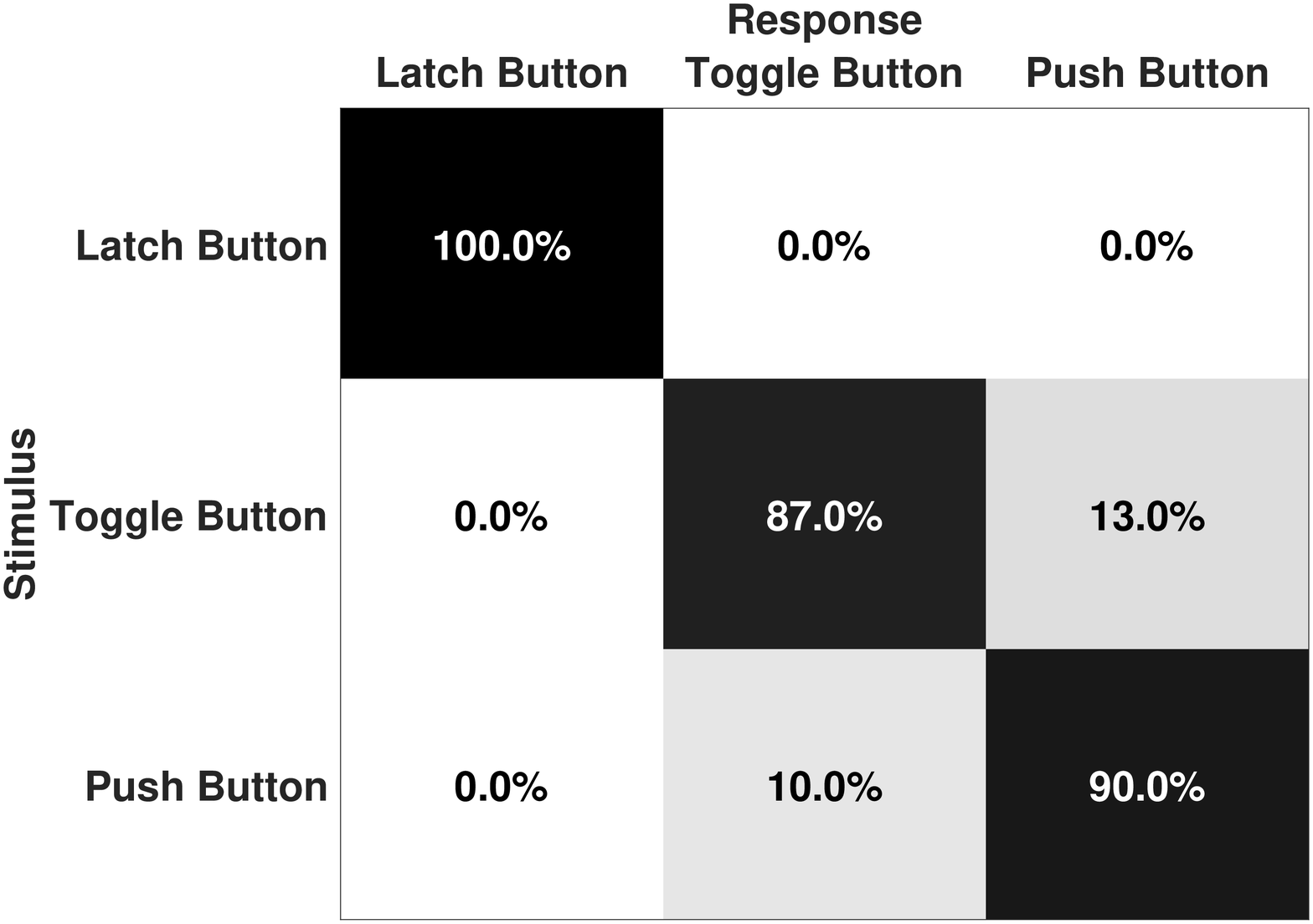}
		
	\end{subfigure}
	\caption{Confusion matrices for  the responses of participants in (a) Group-I with no confirmation (Day-I), (b) Group-I with confirmation (Day-II), (c) Group-II with confirmation (Day-I), and (d) Group-II with no confirmation (Day-II).}\label{fig13}
\end{figure*}

Fig. \ref{fig13} shows the confusion matrices for the responses of both groups on alternate days. The diagonal entries in these matrices show the correct recognition rate in percentage for each button, known as true positive $(TP)$ values whereas the off-diagonal entries are classified as errors. The total number of false negatives $(FN)$ for a button is the sum of the values in the corresponding rows, excluding $(TP)$. The total number of false positives $(FP)$ for a button is the sum of the values of the corresponding columns excluding  $(TP)$. The total number of true negatives $(TN)$ for a button is the sum of all columns and rows excluding the column and row of that button. 

The following performance metrics were used to analyze the results (\citealp{landis1977measurement, fawcett2006introduction, sokolova2009systematic}):
\begin{enumerate}[(a)]
	\item Accuracy (ACC) is calculated as sum of the correct classifications  divided by the total number of classifications (Equation \ref{eq3}).
	\begin{equation}
	ACC = \frac{TP + TN}{Total}\label{eq3}
	\end{equation}
	\item Precision (PREC) is calculated as the number of correct classifications divided by the total number of positive predictions for a specific button (Equation \ref{eq4}). 
	\begin{equation}
	PREC = \frac{TP}{TP+FP}\label{eq4}
	\end{equation}
	\item Sensitivity (SN), also called recall, is calculated as the number of correct positive predictions divided by the total number of stimuli presented for a specific button (Equation \ref{eq5}).
	\begin{equation}
	SN = \frac{TP}{TP+FN}\label{eq5}
	\end{equation}
\end{enumerate}

The sensitivity per button varied from 68\% (\ac{b3}) to 75\% (\ac{b1}) for Group-I on Day-I, (no  confirmation provided during training on digital buttons, see Fig. \ref{fig13a}). After providing confirmation to Group-I during their training on Day-II for digital buttons, the sensitivity was improved by 11\% for \ac{b3}, 16\% for \ac{b2}, and 17\% for \ac{b1}  (see Fig. \ref{fig13b}). The performance metrics in Table \ref{tab3} confirm that the accuracy of identifying \ac{b2} was improved by 12.34\% followed by \ac{b1} (11.33\%) and \ac{b3} (5.67\%) for Group-I on Day-II, when confirmation was provided during training on digital buttons.

\begin{table*}[!htbp]
	\centering
	\caption{Comparison of performance metrics of Group-I for Day-I and Day-II}
	\label{tab3}
	\renewcommand{\arraystretch}{2}
	\begin{tabular}{|c|c|c|c|c|c|c|}
		\hline
		\multirow{2}{*}{\textbf{Metrics}}& \multicolumn{3}{c|}{\textbf{Group-I (Day-I)}}                                                            & \multicolumn{3}{c|}{\textbf{Group-I (Day-II)}}                                                           \\ \cline{2-7}
		& \multicolumn{1}{L{1.9cm}|}{\ac{b1}} & \multicolumn{1}{L{2.0cm}|}{\ac{b2}} & \multicolumn{1}{L{1.9cm}|}{\ac{b3}} & \multicolumn{1}{L{1.9cm}|}{\ac{b1}} & \multicolumn{1}{L{2.0cm}|}{\ac{b2}} & \multicolumn{1}{L{1.9cm}|}{\ac{b3}} \\ \hline
		\textbf{ACC (\%)}   &  82.67\%                             & 76.33\%                             & 81.00\%                            & 94.00\%                             & 88.67\%                             & 86.67\%                            \\ \hline
		\textbf{PREC (\%)} & 73.53\%                          & 63.81\%                          & 73.12\%                          & 90.20\%                          & 83.0\%                           & 80.61\%                         \\ \hline
		\textbf{SN (\%)}   & 75\%                             & 67\%                             & 68\%                            & 92\%                             & 83\%                             & 79\%                            \\ \hline
	\end{tabular}
	\renewcommand{\arraystretch}{1}
\end{table*}

The sensitivity scores of participants in Group-II varied from 81\% (\ac{b3}) to 95\% (\ac{b1}) on Day-I (confirmation provided during training, see Fig. \ref{fig13c}). Sensitivity was improved by 9\% for \ac{b3}, 4\% for \ac{b2}, and 5\% for \ac{b1} on Day-II when Group-II was not provided with any confirmation during their training on digital buttons (see Fig. \ref{fig13d}). It is further confirmed from the performance metrics in Table \ref{tab4} that Group-II could identify \ac{b1} with 100\% precision followed by \ac{b2} (89.69\%) and \ac{b3} (87.38\%) on Day-II when no confirmation was provided during training on digital buttons.

\begin{table*}[h!]
	\centering
	\caption{Comparison of performance metrics of Group-II for Day-I and Day-II}
	\label{tab4}
	\renewcommand{\arraystretch}{2}
	\begin{tabular}{|c|c|c|c|c|c|c|}
		\hline
		\multirow{2}{*}{\textbf{Metrics}}	& \multicolumn{3}{c|}{\textbf{Group-II (Day-I)}}                                                           & \multicolumn{3}{c|}{\textbf{Group-II (Day-II)}}                                                          \\ \cline{2-7}
		& \multicolumn{1}{L{1.9cm}|}{\ac{b1}} & \multicolumn{1}{L{2.0cm}|}{\ac{b2}} & \multicolumn{1}{L{1.9cm}|}{\ac{b3}} & \multicolumn{1}{L{1.9cm}|}{\ac{b1}} & \multicolumn{1}{L{2.0cm}|}{\ac{b2}} & \multicolumn{1}{L{1.9cm}|}{\ac{b3}} \\ \hline
		\textbf{ACC (\%)}   &  97.00\%                             & 87.33\%                             & 88.33\%                            & 100\%                             & 92.33\%                             & 92.33\%                            \\ \hline
		\textbf{PREC (\%)} & 95.96\%                          & 79.81\%                          & 83.51\%                         & 100\%                            & 89.69\%                          & 87.38\%                         \\ \hline
		\textbf{SN (\%)}   & 95\%                             & 83\%                             & 81\%                            & 100\%                             & 87\%                             & 90\%                            \\ \hline
	\end{tabular}
	\renewcommand{\arraystretch}{1}
\end{table*}

Fig. \ref{fig13} shows that if confirmation was provided to the participants during training on the digital buttons, their performance was improved especially in the case of \ac{b2} and \ac{b3}. However, we observed that both groups found \ac{b1} distinct as compared to \ac{b2} and \ac{b3}, but confused \ac{b2} with \ac{b3} irrespective of the confirmation provided. 

We further performed a three-way ANOVA on the performance metrics with two within-subjects factors (confirmation and button-type; latch, toggle, and push button) and one between-subjects factor (group number). First, we performed the Mauchly's test of sphericity to check whether  the differences between the levels of the within-subject factors have equal variance. If the sphericity assumption was violated, the degrees of freedom were corrected using Greenhouse-Geisser  correction. Then, Bonferroni corrected post-hoc analysis was carried out to investigate where the statistically significant differences between the levels of within-subject factors lie.

Providing confirmation to the participants during training has a significant main effect on all the performance metrics $(p < 0.01)$. When the confirmation was provided, the performance metrics were improved. The button-type also has a statistically significant main effect on all of the performance metrics $(p < 0.01)$. Compared to \textit{Toggle} and \textit{Push Button}, \textit{Latch Button} had the highest accuracy, precision, and sensitivity scores.    

There was a significant interaction between confirmation and group number ($p < 0.01$), suggesting that, providing confirmation to Group-II first and Group-I later during training session yields a different effect on  the performance metrics. On Day-I, Group-II performed better than Group-I, since they received confirmation. The interaction between the button-type and group number was not significant. The interaction effect between confirmation and button-type on response was also non-significant.

Our post-hoc analysis showed that participants were significantly better at relating digital \ac{b1} with its physical counterpart in comparison with  \ac{b2} and \ac{b3}  $(p<0.001)$. However, we could not find significant differences in relating \ac{b2} and \ac{b3} to their physical counterparts irrespective of the confirmation provided during the training session.   

\subsection{Experiment-II}
We applied a normalization to the subjective ratings based on the method suggested by \cite{murray2003psychophysical}. For this purpose, we first computed the geometric mean of all responses, called Grand Geometric Mean ($GGM$), and the geometric mean of each participant ($GM_P$). We then obtained the normalized value for each participant by $GGM/GM_P$.

\begin{figure}[!htp]
	%\fulltextwidth
	\centering
	\begin{subfigure}[b]{0.5\textwidth}
		\centering
		\caption{}	
		\includegraphics[width=1\linewidth]{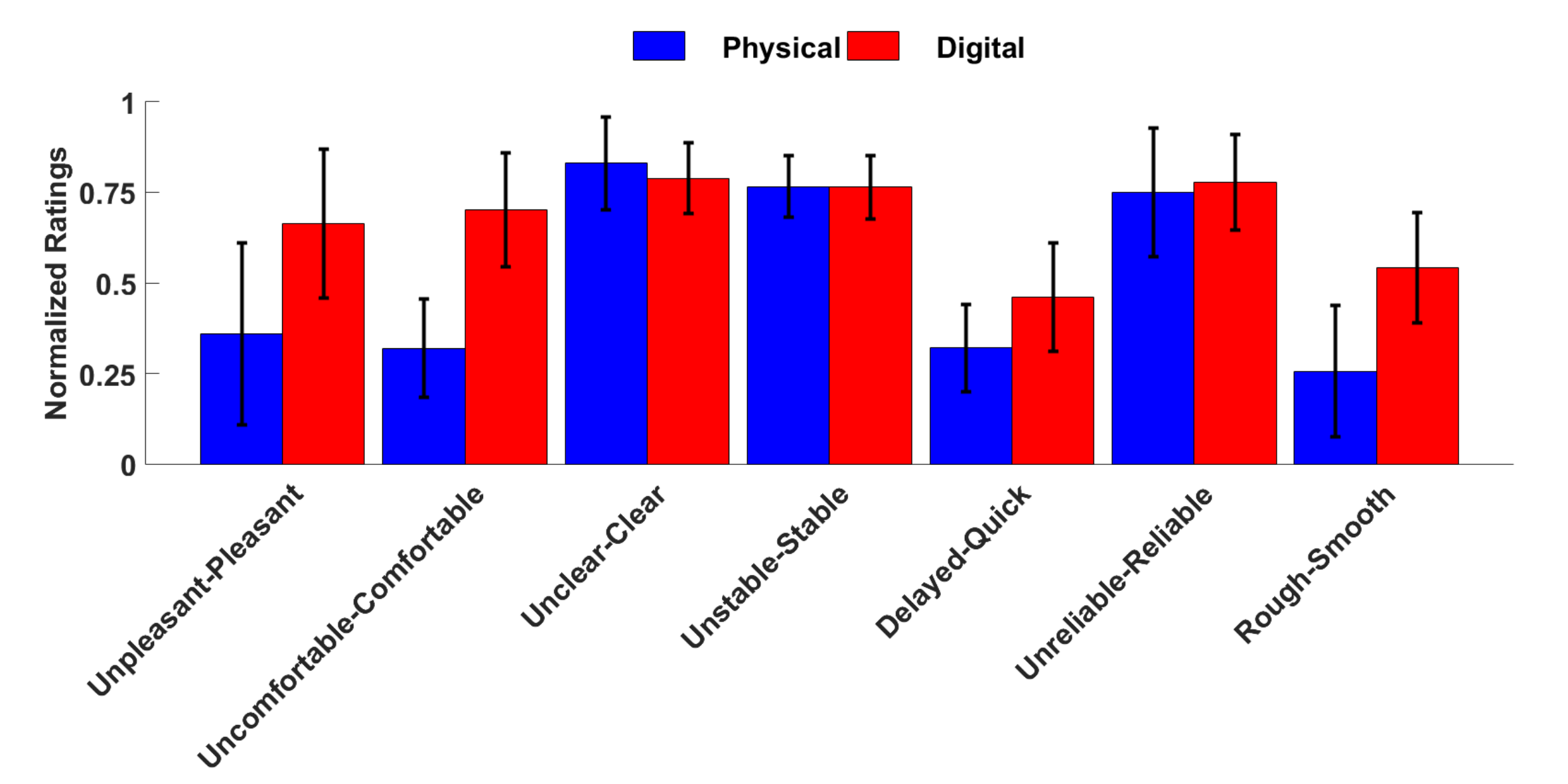}
		
		\label{fig14a}
	\end{subfigure}%
	%	\vspace{0.5} 
	\vspace{1em}
	\begin{subfigure}[b]{0.5\textwidth}
		\centering
		\caption{} 
		\includegraphics[width=1\linewidth]{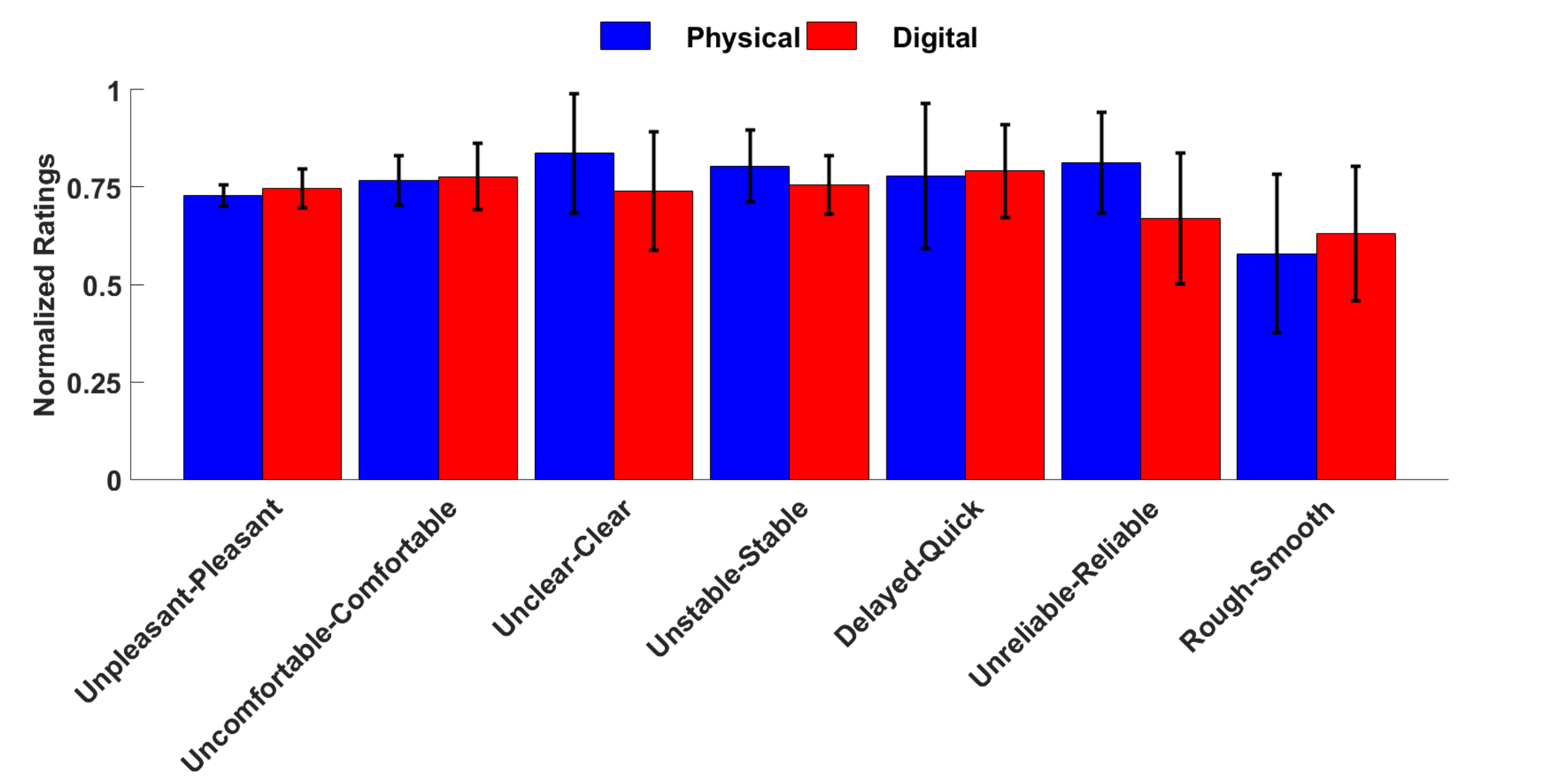}
		
		\label{fig14b}
	\end{subfigure}
	
	\begin{subfigure}[b]{0.5\textwidth}
		\centering
		\caption{}
		\includegraphics[width=1\linewidth]{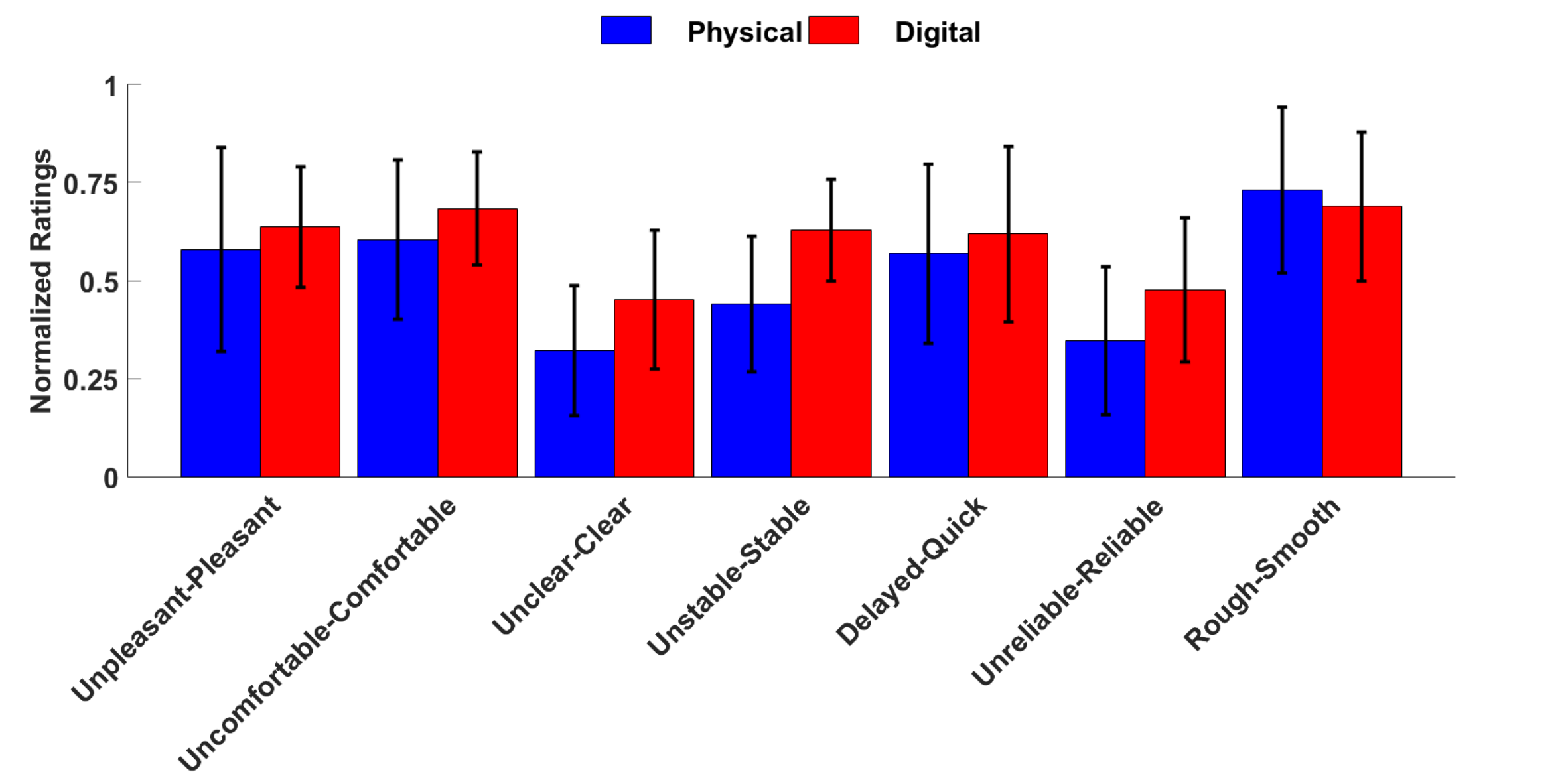}
		
		\label{fig14c}
	\end{subfigure}%
	\caption{Normalized adjective ratings of all three physical and digital buttons with their standard deviations: (a) \ac{b1}, (b) \ac{b2}, and (c) \ac{b3}.}\label{fig14}
\end{figure}

The average ratings obtained in Experiment-II and their standard deviations are shown in Fig. \ref{fig14} for each button. Participants have rated digital \ac{b1} similar to physical \ac{b1} on \textit{Unstable-Stable}, \textit{Unclear-Clear} and \textit{Unreliable-Reliable} adjective pairs. However, they felt digital \ac{b1} more pleasant, comfortable, quick, and smooth compared to its physical counterpart. Participants have felt digital \ac{b2} similar to its physical counterpart for almost all adjective pairs except for \textit{Unreliable-Reliable}. According to participants' ratings, digital \ac{b2} is less reliable than its physical counterpart. They felt digital \ac{b3} similar to its physical counterpart for \textit{Rough-Smooth} adjective pair.
We conducted a two-way multivariate analysis of variance (MANOVA) as described by \cite{meyers2016applied} with two independent variables (button-type, button-category; physical and digital) and seven dependent variables (seven adjective pairs listed in Table \ref{tab2}).  

The MANOVA found significant main effects of both button-category ($F(7, 108) = 7.46, p < 0.001, \eta^2=0.326$), and button-type ($F(14, 218) = 26.41, p < 0.001,  \eta^2=0.63$) on the combined dependent variables. The interaction between button-category and button-type was also significant ($F(14, 218) = 5.36, p < 0.001, \eta^2=0.26$). Individual ANOVA found a non-significant main effect of button-category for two adjective pairs: \textit{Unclear-Clear}  and \textit{Unreliable-Reliable}, showing that participants have rated digital buttons similar to their physical counterparts for these adjective pairs. Our post-hoc analysis showed that participants rated two of the  three digital buttons (\ac{b2} and \ac{b3}) similar to their physical counterparts for four adjective pairs: \textit{Unpleasant-Pleasant, Uncomfortable-Comfortable,  Delayed-Quick,} and \textit{Rough-Smooth}. Digital \ac{b1} was rated similar to its physical counterpart for three adjective pairs: \textit{Unclear-Clear, Unstable-Stable,} and \textit{Unreliable-Reliable}. 

\section{Discussion}
In this study, we  recorded force, acceleration, and voltage data for activation state from three physical buttons to identify their distinct response characteristics. We used the acceleration profile recorded for each physical button to generate the vibrotactile stimulus displayed for its digital counterpart on the interaction surface of HapTable. To find the most optimal signal for each button, a dynamic time warping algorithm was applied to the recorded acceleration profile first. After obtaining the optimal signals for each button, we mapped the acceleration profile to the actuation voltage signal using the transfer function of the interaction surface for the point where the digital buttons were displayed to the users. Since it was not possible to display force feedback to the users in our system, the information gained from the recorded force data  was associated with the finger contact area of each participant pressing the digital button to determine the instant of its activation. 

We conducted a user study in which we asked the participants to match the vibrotactile stimulus, displayed to them through the touch surface, with one of the three physical buttons that they had interacted with. Hence, the chance level of the experiment was 33\%. The correct recognition rate of both groups was significantly higher than the chance level for each button (Tables \ref{tab3} and \ref{tab4}).  

Our experimental results revealed that both groups had confusion in matching \ac{b2} and \ac{b3} to their physical counterparts irrespective of the availability of the confirmation during training step. Six out of ten participants in Group-I on Day-I perfectly matched the digital \ac{b1} to its physical counterpart, whereas three participants confused \ac{b2} with \ac{b3} and vice versa. Only, one participant perfectly matched all digital buttons to their physical counterparts for all trials. Note that there was no confirmation provided to Group-I on Day-I. After providing confirmation during training, the overall performance was improved by 15\%. However, the \ac{b2} was still confused with \ac{b3}. Eight participants from Group-II on Day-I could match digital \ac{b1} to its physical counterpart with 100\% accuracy. Two out of eight participants perfectly matched digital buttons with their physical counterparts and three participants achieved 99\% accuracy and confused \ac{b2} with \ac{b3} only once. Rest of the participants confused \ac{b1} with \ac{b2} and \ac{b2} with \ac{b3}. The overall accuracy was 86.33\% for Group-II on Day-I and improved by 6\% on Day-II when no confirmation was provided during training. Overall, the participants correctly matched the given digital buttons with their physical counterparts with an accuracy of 83\%. 

After the matching experiment, we verbally asked participants to comment on their subjective experience with digital buttons. Majority of the participants were confident that they matched the \ac{b1} correctly. Some participants stated that the duration of haptic stimuli (vibration duration) helped them to distinguish the \ac{b2} from \ac{b3}, whereas some stated that they made a decision based on the pressure they applied to the touchscreen to activate the button, which is related to the finger contact area. Fifteen out of twenty participants preferred \ac{b2} and \ac{b3} over \ac{b1}. These results are in line with some previous studies by \cite{c4, c10} and \cite{c8}. They emphasized the importance of keeping duration short for button click vibrations. Furthermore, \cite{shin2014effect} also highlighted that tactile stimulus responding to pressing a digital button on touch surface with a rapid response time provides a realistic feeling of physically clicking a button. This is also confirmed in our experimental results. Participants have preferred digital buttons with short vibration duration (\ac{b2} and \ac{b3}) as compared to the one with longer vibration duration (\ac{b1} in our case). In summary, participants reported rich and diverse haptic sensations for the digital buttons displayed in our study. 

We also asked participants to rate their subjective tactile feelings of physical and digital buttons using seven adjective pairs (Table \ref{tab2}). Our statistical analysis revealed that participants have rated two out of three digital buttons (\ac{b2} and \ac{b3}) similar to their physical counterparts for four adjective pairs: \textit{Unpleasant-Pleasant, Uncomfortable-Comfortable,  Delayed-Quick,} and \textit{Rough-Smooth}. Furthermore, our results also showed that participants perceived digital buttons more pleasant, comfortable, and smooth as compared to their physical counterparts.

\section{Conclusion}
Although pressing and turning on/off a button is a simple physical interaction in our daily life, imitating it in a digital world, especially on touchscreens, appears to be more difficult. In fact, this is not surprising if we look at this simple physical interaction more carefully. First, our experimental data of acceleration and force quickly reveals that each button has distinct response characteristics (see Fig. \ref{fig6}). Second, the identification of sub-events and their timing are important (for example, the \ac{b1} in our study is activated in two stages with a certain time interval between them, see Fig. \ref{fig6}). Third, the instant of button activation with respect to the acceleration profile is critical for proper rendering of digital buttons on touchscreens (again, for example, \ac{b1} is activated quite late compared to others, see Fig. \ref{fig6}). When humans interact with the physical buttons in the real world, they receive kinesthetic haptic feedback, which is not possible yet to replicate in the digital world on a touchscreen. Currently, only tactile haptic feedback can be displayed through a touchscreen in the form of vibrotactile stimulus as it is done in this study or friction modulation via ultrasonic and electrostatic actuation as in our other studies (\citealp{saleem2018psychophysical, c17, vardar2018tactile}). Hence, for example, it is highly difficult to convey the springiness of a physical button via tactile feedback on a touchscreen (\ac{b3} in our study is a good example of this case). Moreover, humans are good at controlling force in their daily interactions with physical objects, and buttons are no exception for this. For example, we know exactly how much pressure to apply to turn on/off a light switch in the real world. In particular, the accumulated pressure is related to the activation and deactivation instant and period of a button. It is also important to note here that there are also differences between the magnitude of force/pressure applied by the individuals (see Fig. \ref{fig7}). Since it is highly difficult to measure contact forces in systems utilizing a touch surface, the earlier studies have mostly ignored the role of accumulated force/pressure in a tactile rendering of digital buttons. In our study, we relate finger contact force with the contact area and activate a digital button based on the instantaneous finger contact area of the participant, which is measured by an infrared camera during real-time interaction. The digital button is activated when the participant applies sufficient pressure to the touchscreen with her/his finger such that the instantaneous contact area exceeds the pre-determined threshold value. Even though we tried to imitate the feelings of physical buttons in the form of mechanical vibration, as our results show that participants were able to match digital buttons with their physical counterparts on only four adjective pairs. This might be related to the kinesthetic feedback, size, and texture of the physical buttons. Also, we were not able to stimulate all the mechanoreceptors that were stimulated when interacting with the physical buttons. For example, (Slow Adapting) SA-I is responsible for detection of edges, intensity, and pressure which was not stimulated when interacting with the digital buttons. \textcolor{blue}{In fact, Fig. \ref{fig9} shows that the frequency spectrum of the recorded and reconstructed acceleration signals are quite different due to the frequency modulation, though their profiles are similar in time domain. We modulated the carrier signal with the resonance frequency of the table (263.5 Hz) to make the vibrations more salient for the user, but this obviously changed the frequency content of the reconstructed acceleration signal and made it more “vibratory” compared to the acceleration signal recorded from the physical buttons.} In the future, we will focus on the haptic display of other digital widgets such as slider and knob on touch surfaces using the proposed data-driven technique. Since our HapTable enables to modulate friction on the touch surface via electrostatic actuation, it will be possible to display multi-modal tactile feedback (vibration in the normal direction and friction force in the tangential direction) for rendering digital slider and knob.

\section*{References}
\bibliographystyle{elsarticle-harv}%\biboptions{authoryear}
\bibliography{MyRef}

\end{document}

%% file: Acronym.tex
\DeclareAcronym{b1}{
	short = \textit{Latch Button},
	long = Button \# 1, 
class = abbrev
}
\DeclareAcronym{b2}{
	short = \textit{Toggle Button},
	long = Button \# 2, 
	class = abbrev
}
\DeclareAcronym{b3}{
	short = \textit{Push Button},
	long = Button \# 3, 
	class = abbrev
}